\documentclass{iopart}

\usepackage{graphicx,epsfig,sidecap}
\usepackage{bm}

\catcode`@=11 
\renewcommand\tableofcontents{%
  \section*{\contentsname}%
  \@starttoc{toc}%
}
\catcode`@=12

\def\be{\begin{equation}}
\def\ee{\end{equation}}

\def\bea{\begin{eqnarray}}
\def\eea{\end{eqnarray}}

\def\Tr{{\rm Tr}\,}
\def\HH{\mathcal H}
\def\e{\epsilon}

\def\l{\lambda}
\def\lm{{\l_{\rm max}}}

\newcommand{\rx}{{ x}}
\newcommand{\ry}{{\tau}}
\def\lt#1{\left#1}
\def\rt#1{\right#1}
\def\t#1{\tilde{#1}}

\newcommand{\C}{{\mathbf{C}}}

\newcommand{\bra}{\langle}
\newcommand{\ket}{\rangle}
\newcommand{\tw}{{\cal T}}

\begin{document}

\title[Entanglement entropy and CFT]
{Entanglement entropy and conformal field theory}

\author{Pasquale Calabrese$^1$ and John Cardy$^{2}$}
\address{$^1$Dipartimento di Fisica dell'Universit\`a di Pisa and INFN,
             Pisa, Italy.\\
         $^2$ Oxford University, Rudolf Peierls Centre for
          Theoretical Physics, 1 Keble Road, Oxford, OX1 3NP, United Kingdom
          and All Souls College, Oxford.}

\date{\today}

\begin{abstract}
We review the conformal field theory approach to entanglement
entropy in 1+1 dimensions. 
We show how  to apply these methods to the calculation of
the entanglement entropy of a single interval, and the
generalization to different situations such as finite size,
systems with boundaries, and the case of several disjoint
intervals. We discuss the behaviour away from the critical point
and the spectrum of the reduced density matrix. Quantum quenches,
as paradigms of non-equilibrium situations, are also considered.

\end{abstract}

\maketitle

\tableofcontents

\section{Introduction}

Entanglement is one of the most fundamental and fascinating features of 
quantum mechanics, yet in some ways the most mysterious: performing 
a local measure may instantaneously affect the outcome of local measurements 
far away. 
This phenomenon has been the basis for the development of such new branches 
of research as quantum information and communication. 
A very recent and rich field of research concerns the understanding of 
the role of entanglement in many-body systems.

Beside its own fundamental theoretical interest, a
principal reason for the success of the entanglement entropy as
an entanglement measure in extended quantum systems is surely its
universal scaling at one-dimensional (1D) conformal critical
points. The equation \cite{Holzhey,Vidal,cc-04} 
\be S_A=
\frac{c}3\ln \frac{\ell}a +c'_1\,, \label{SA1} 
\ee 
has become one of the most ubiquitous formulas in the last five years' 
literature, appearing in fields as apparently unrelated as quantum
information, condensed matter, and high energy physics. 
The reasons for this prominence are clear: it is a single quantity,
easily measurable in numerical simulations, that at the same time
gives the location of the critical point and one of its most
important universal signatures, the {\it central charge} $c$ of
the underlying conformal field theory (CFT).

The aim of this review is to give a self-contained presentation of
most of the results for the entanglement entropy that can be
obtained by means of CFT. Some other important features of the
entanglement in extended quantum systems will only be considered
on passing. For a comprehensive treatment of these aspects we
refer the reader to the other reviews in this special issue
\cite{ccd-sp} and to the already existing ones \cite{af-rev,e-rev}.

The plan of this article is as follows. In Sec.~\ref{sec2} we
review the CFT (and more generally quantum field theory) 
approach to entanglement entropy based on the
replica trick, the mapping to the partition function on Riemann
surfaces, and the introduction of twist fields. In Sec.~\ref{sec3}
we apply these methods to the calculation of the entanglement
entropy of a single interval, showing in particular how to obtain
Eq.~(\ref{SA1}), and generalizing it to several different
situations, like finite size, finite temperature, systems with
boundaries and defects. In Sec.~\ref{sec4} we consider the case of
several disjoint intervals, with particular attention to the case
of two intervals, where several results are now available. In
Sec.~\ref{sec5} we consider massive perturbations to the conformal
behaviour in the regime when the mass is small and the systems
still retain signatures of the close conformal critical point. In
Sec.~\ref{sec6} we derive the consequences of the conformal
scaling for the full spectrum of the reduced density matrix. In
Sec.~\ref{sec7} we discuss the CFT approach to non-equilibrium
situations known as quantum quenches. Finally in Sec.~\ref{sec8}
we report on an interesting proposal of Klich and Levitov
\cite{kl-08} to measure the entanglement entropy in real experiments.

\section{Entanglement, replicas, Riemann surfaces, twist fields and all that}
\label{sec2}

\subsection{Basic definitions}

Let $\rho$ be the density matrix of a system, which we take to be
in the pure quantum state $|\Psi\rangle$, so that
$\rho=|\Psi\rangle\langle\Psi|$. Let the Hilbert space be written
as a direct product $\HH=\HH_A\otimes\HH_B$. $A$'s reduced density
matrix is $\rho_A=\Tr_B \rho$. The entanglement entropy is the
corresponding von Neumann entropy \be S_A=-\Tr\, \rho_A \ln
\rho_A\,, \label{Sdef} \ee and analogously for $S_B$. When $\rho$
corresponds to a pure quantum state $S_A=S_B$. For future use, we
also define the R\'enyi entropies 
\be S^{(n)}_A= \frac{1}{1-n} \ln {\rm Tr}\,\rho_A^n\,, \ee 
that also are characterized by
$S_A^{(n)}=S_B^{(n)}$ whenever $\rho$ corresponds to a pure
quantum state. From these definitions $S_A=\lim_{n\to1}
S_A^{(n)}$.

When a system is in a mixed state the entanglement entropy is not
any longer a good measure of entanglement since it clearly mixes
quantum and classical correlations (e.g. in an high temperature
mixed state, it must reproduce the extensive result for the
thermal entropy that has nothing to do with entanglement.) This is
also evident from the fact that $S_A$ is no longer equal to $S_B$.
A quantity that is easily constructed from the knowledge of $S_A$
and $S_B$ is the so called mutual information, defined from the
R\'enyi entropy as \be
I^{(n)}_{A:B}=S^{(n)}_A+S^{(n)}_B-S^{(n)}_{A\cup B}\,, \ee that is
by definition symmetric in $A$ and $B$. $I^{(n)}_{A:B}$ has {\it
not} all the correct properties to be an entanglement measure (see
the review by Amico and Fazio in this volume \cite{af-rev2} for a
discussion of various entanglement measures), but it has the
fundamental property of satisfying the area law \cite{s-93,e-rev}
even at finite temperature \cite{wvhc-08}.

\subsection{A replica approach}

When dealing with a statistical model with a finite number of
degrees of freedom, the most direct way to obtain the entanglement
entropy is to construct the reduced density matrix (or at least
its eigenvalues $\l_i$ as in DMRG) exactly or numerically and
then, by brute force or by analytic methods, calculate the sum
$S_A=-\sum\l_i \ln \l_i$. Several examples of how this can be
worked out, even analytically, for the simplest models are
reported in other reviews in this special issue. However,
calculating the full reduced density matrix for a generic
interacting quantum field theory remains a daunting challenge from
an operational point of view, and so here we will take a different
route, following our previous papers
\cite{cc-04,cc-05p,ccd-07,c-SF}. The approach is reminiscent of
the ``replica trick'' in disordered systems and was present in an
embryonic form also in the early paper by Holzhey, Larsen, and
Wilczek \cite{Holzhey}. Let us start by considering a lattice
model. The eigenvalues of the reduced density matrix $\l_i$ lie in
the interval $[0,1]$ and $\sum\l_i=1$. Thus, for any $n\geq1$
(even not integer), the sum ${\rm Tr}\,\rho_A^n=\sum_i\l_i^n$ is
absolutely convergent and therefore analytic for all ${\rm
Re}\,n>1$. The derivative wrt $n$ therefore also exists and is
analytic in the region. Moreover, if the entropy
$S_A=-\sum_i\l_i\log\l_i$ is finite, the limit as $n\to1^+$ of the
first derivative converges to this value. Thus if we are able to
calculate ${\rm Tr}\,\rho_A^n$ for any $n\geq1$ we have the
entanglement entropy as
\begin{equation}
S_A=-\lim_{n\to1}{\partial\over\partial n}{\rm Tr}\,\rho_A^n
= 
\lim_{n\to1} S^{(n)}_A\,.
\label{SA3}
\end{equation}
However, calculating ${\rm Tr}\,\rho_A^n$ for a generic real $n$
in a quantum field theory is still an hopeless task. And here the
``replica trick'' enters: we compute ${\rm Tr}\,\rho_A^n$ only for
positive integral $n$ and then analytically continue it to a
general complex value. We will see that the calculation of ${\rm
Tr}\,\rho_A^n$ reduces to that of a partition function on a
complicated Riemann surface (or equivalently to the correlation
function of specific {\it twist fields}) that is analytically
achievable in a quantum field theory. The problem is then moved to
the existence and the uniqueness of a proper analytic continuation
to extract the entanglement entropy (as in any approach based on
replicas). In some cases this is straightforward, in others
difficult, and in several others beyond our present understanding.

\subsection{Path integral formulation and Riemann surfaces}

\begin{figure}
\includegraphics[width=0.32\textwidth]{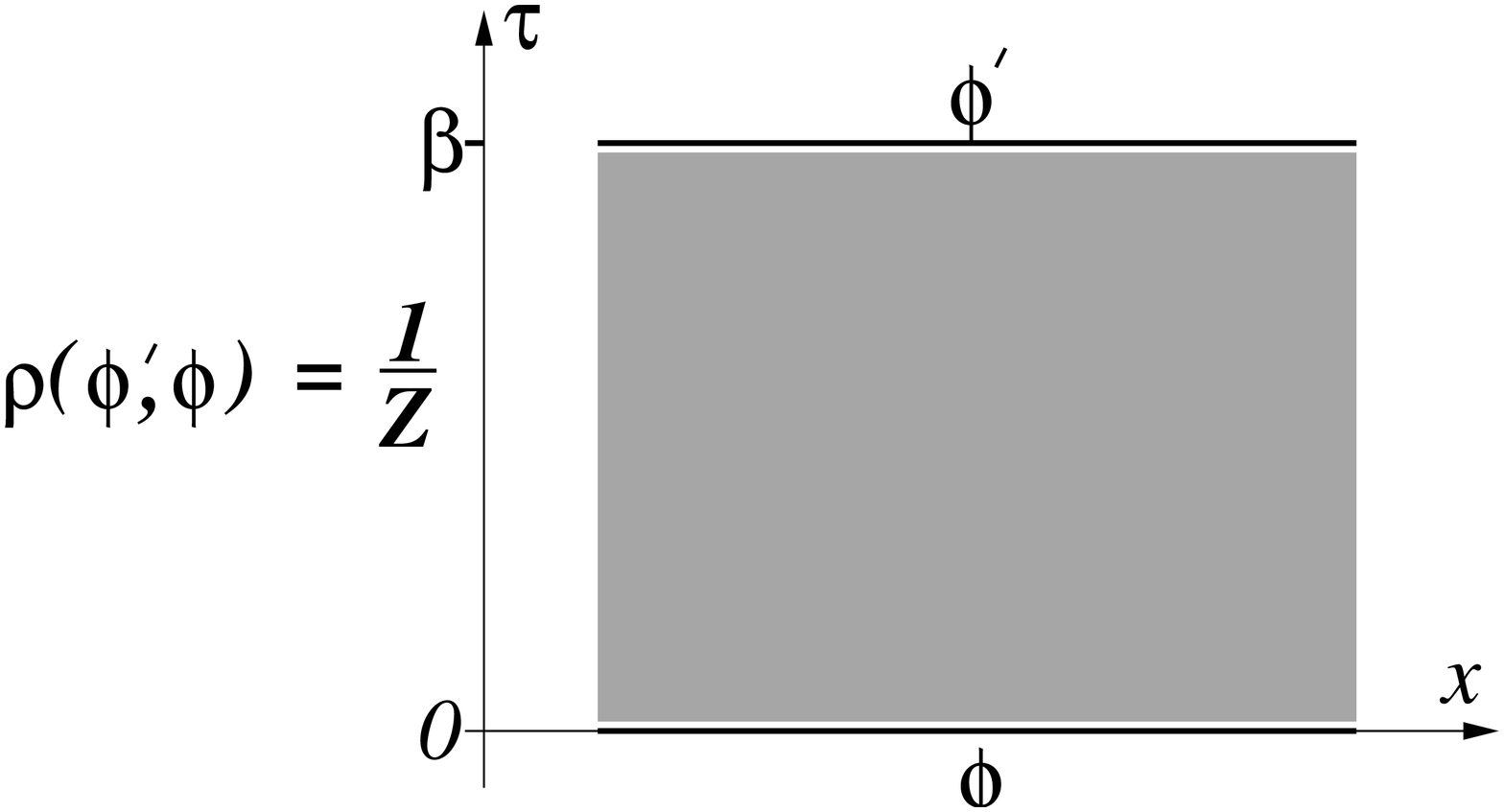}
\includegraphics[width=0.33\textwidth]{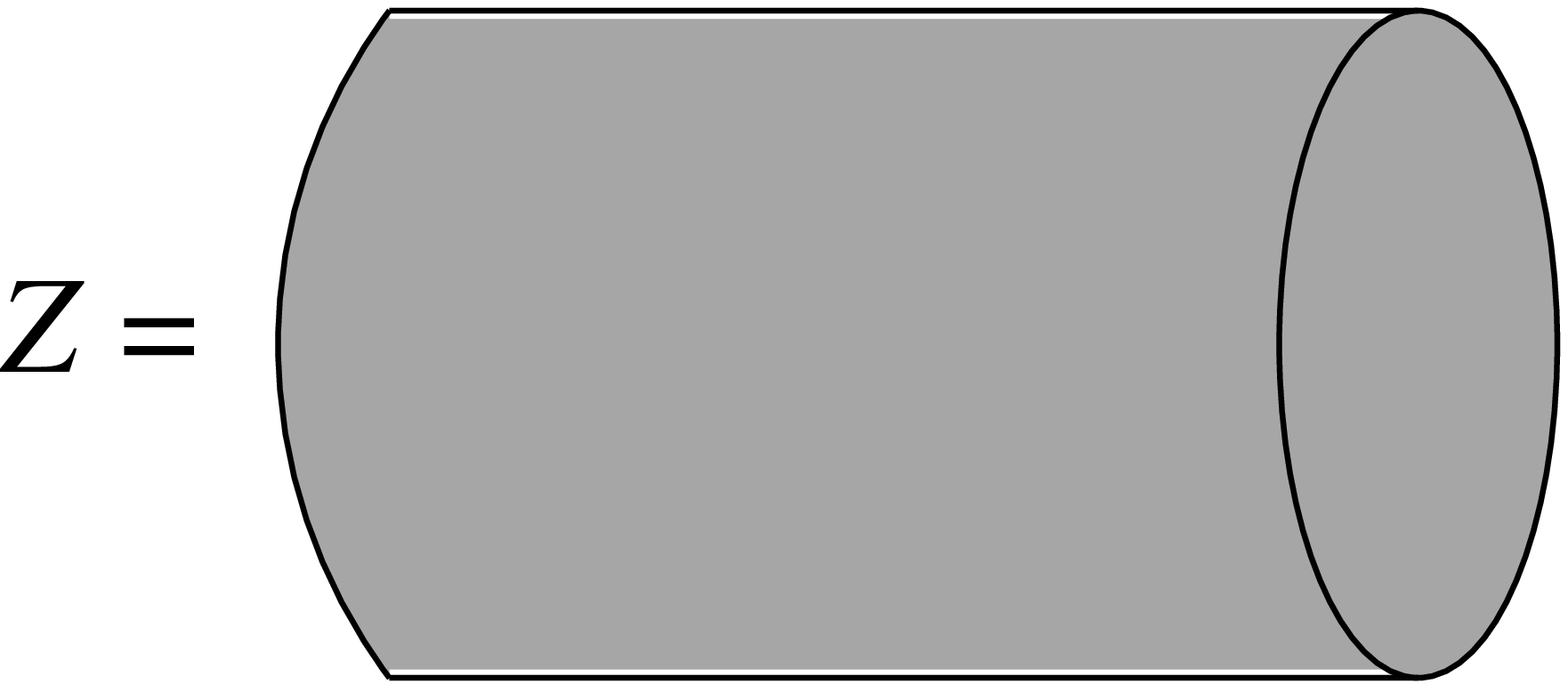}\hspace{0.03\textwidth}
\includegraphics[width=0.30\textwidth]{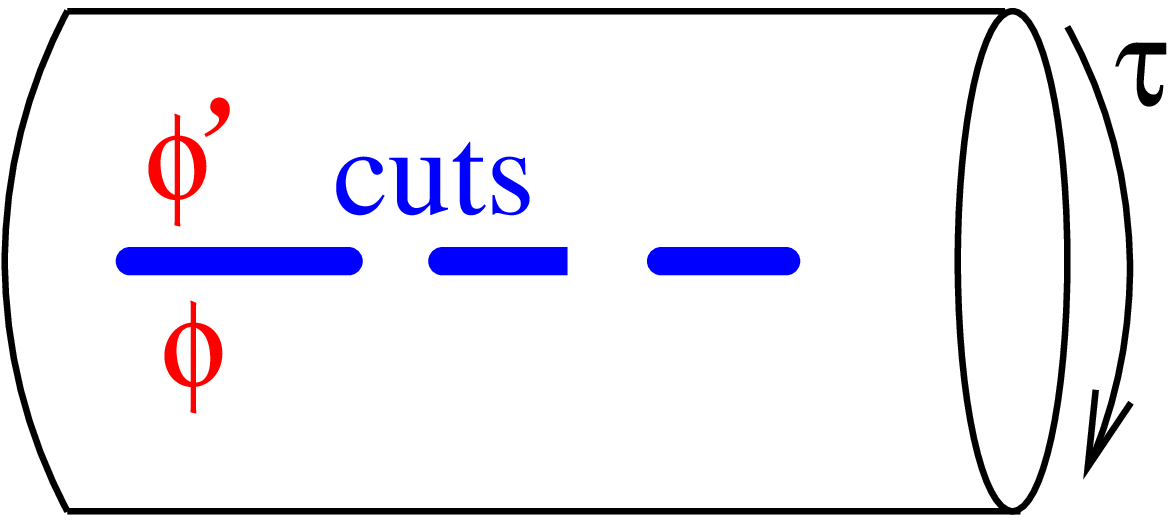}
\caption{From density matrix to reduced density matrix.
Left: Path integral representation of $\rho(\phi|\phi')$.
Center: The partition function $Z$ is obtained by sewing together the edges
along $\tau=0$ and $\tau=\beta$ to form a cylinder of circumference $\beta$.
Right: The reduced density matrix $\rho_A$ is obtained by sewing together
only those points $x$ which are not in $A$.
}
\label{cil}
\end{figure}

To begin with a well-defined problem, let us consider a lattice
quantum theory in one space and one time dimension (the
generalization to higher spatial dimensions is straightforward).
The lattice sites are labelled by a discrete variable $x$ and the
lattice spacing is $a$. The domain of $x$ can be finite, i.e. some
interval of length $L$, semi-infinite, or infinite. Time is
continuous. A complete set of local commuting observables will be
denoted by $\{\hat\phi_x\}$, and their eigenvalues and
corresponding eigenstates by $\{\phi_x\}$ and $|\{\phi_x\}\rangle$
respectively. (For a bosonic lattice field theory, these will be
the fundamental bosonic fields of the theory; for a spin model
some particular component of the local spin.) The states
$\otimes_x|\{\phi_x\}\rangle=|\prod_x\{\phi_x\}\rangle$ form a
basis. The dynamics of the theory is described by the Hamiltonian
$H$. The elements of the density matrix $\rho$ in a thermal state
at inverse temperature $\beta$ are
\begin{equation}
\fl\rho(\{\phi_x\}|\{\phi'_{x'}\})\equiv
\langle \prod_x\{\phi_x\}|\rho |\prod_{x'}\{\phi'_{x'}\}\rangle
=Z(\beta)^{-1}
\langle \prod_x\{\phi_x\}|e^{-\beta H}|\prod_{x'}\{\phi_{x'}\}\rangle\,,
\end{equation}
where $Z(\beta)={\rm Tr}\,e^{-\beta H}$ is the partition function.
This may be written as a path integral on the imaginary time
interval $(0,\beta)$:
\begin{equation}
\label{pathi}
\fl \rho(\{\phi_x\}|\{\phi'_{x'}\})=
Z^{-1}\int[d\phi(y,\tau)]
\prod_{x'}\delta(\phi(y,0)-\phi'_{ x'})\prod_x
\delta(\phi(y,\beta)-\phi_x)\,e^{-S_E}\,,
\end{equation}
where the euclidean action is $S_E=\int_0^\beta  L d\tau$, with
$L$ the euclidean lagrangian. This is illustrated in the left
panel of Fig.~\ref{cil}. Here the rows and columns of the reduced
density matrix are labelled by the values of the fields at
$\tau=0,\beta$.

The normalization factor $Z$ is the partition function, and
ensures that ${\rm Tr}\rho=1$. It is found by setting
$\{\phi_x\}=\{\phi'_{x}\}$ and integrating over these variables.
In the path integral, this has the effect of sewing together the
edges along $\tau=0$ and $\tau=\beta$ to form a cylinder of
circumference $\beta$ as depicted in the center of Fig.~\ref{cil}.

Now let $A$ be a subsystem consisting of the points $x$ in the
disjoint intervals $(u_1,v_1),\ldots,(u_N,v_N)$. An expression for the
the reduced density matrix $\rho_A$ is  obtained from (\ref{pathi})
by sewing together only those points $x$ which are not in $A$. This
has the effect of leaving open cuts, one for each interval
$(u_j,v_j)$, along the line $\tau=0$ as in the right panel of figure \ref{cil}.

\begin{figure}
\includegraphics[width=8cm,height=5cm]{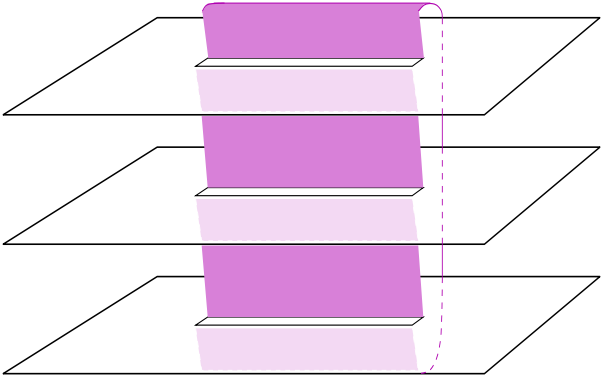}
\caption{A representation of the Riemann surface ${\cal R}_{3,1}$.
Reprinted with permission from \cite{ccd-07}.}
\label{fig-sheets}
\end{figure}

We may then compute ${\rm Tr}\,\rho_A^n$, for any positive integer
$n$, by making $n$ copies of the above, labelled by an integer $j$
with $1\leq j\leq n$, and sewing them together cyclically along
the cuts so that
$\phi_j(x,\tau=\beta^-)=\phi_{j+1}(x,\tau=0^+)$ and
$\phi_n(x,\tau=\beta^-)=\phi_1(x,\tau=0^+)$ for all $x\in A$. This
defines an $n$-sheeted structure depicted for $n=3$ and in the
case when $A$ is a single interval in Fig.~\ref{fig-sheets}. The
partition function on this surface will be  denoted by $Z_n(A)$
and so
\begin{equation}
\label{ZoverZ}
{\rm Tr}\,\rho_A^n={Z_n(A)\over Z^n}\,.
\end{equation}
When the right hand side of the above equation has a unique
analytic continuation to ${\rm Re}\,n>1$, its first derivative
at $n=1$ gives the required entropy
\begin{equation}
S_A=-\lim_{n\to1}{\partial\over\partial n}{Z_n(A)\over
Z^n}\,.
\end{equation}

So far, everything has been for a discrete space domain. We now
discuss the continuum limit, in which $a\to0$ keeping all other
lengths fixed. The points $x$ then assume real values, and the
path integral is over fields $\phi(x,\tau)$ on an {\it $n$-sheeted
Riemann surface}, with branch points at $u_j$ and $v_j$. In this
limit, $S_E$ is supposed to go over into the euclidean action for
a quantum field theory. We indicate these $n$-sheeted surfaces
with ${\cal R}_{n,N}$ and they are fully defined by the $2N$
branch points $u_j$ and $v_j$. Whenever the value of $n$ and $N$
is not important, we will simply indicate the surface with ${\cal
R}$.

In the following, we will restrict our attention to the case when
the quantum field theory is Lorentz invariant, since the full power of
relativistic field theory can then be brought to bear.
The behaviour of partition functions in this limit has been well
studied. In two dimensions, the logarithm of a general partition function
$Z$ in a domain with total area $\cal A$ and with boundaries of total
length $\cal L$ behaves as
\begin{equation}
\label{divs}
\log Z=f_1{\cal A}a^{-2}+f_2{\cal L}a^{-1}+\ldots
\end{equation}
where $f_1$ and $f_2$ are the non-universal bulk and boundary free
energies. Note, however, that these leading terms {\em cancel} in the
ratio of partition functions in (\ref{ZoverZ}).

In a conformal field theory, as was argued by Cardy and
Peschel \cite{CardyPeschel},
there are also {\em universal} terms proportional to $\log a$.  These
arise from points of non-zero curvature of the manifold and its
boundary. In our case, these are conical singularities at the branch
points. In fact, 
it is precisely these logarithmic terms which give rise to the
non-trivial dependence of the final result for the entropy on the
short-distance cut-off $a$. For the moment let us simply remark
that, in order to achieve a finite limit as $a\to0$, the right
hand side of (\ref{ZoverZ}) should be multiplied by some
renormalization constant ${\cal Z}(A,n)$.

\subsection{From replicated world-sheet to replicated target-space:
Twist fields}
\label{Sectwist}

In the simplest instances it is possible to directly calculate the partition
function on a $n$-sheeted Riemann surface, but in most of the cases
this is very difficult.
However, the surface we are dealing with has curvature
zero everywhere except at a finite number of points (i.e. the boundaries
between $A$ and $B$ $u_j,v_j$ above).
Since the lagrangian density does not depend explicitly on the Riemann
surface ${\cal R}$ as a consequence of its locality, it is expected that the
partition function can be expressed as an object calculated from
a model on the complex plane $\C$, where the structure of the Riemann surface is
implemented through appropriate boundary conditions around the
points with non-zero curvature.
Consider for instance the simple Riemann surface ${\cal R}_{n,1}$ needed
for the calculation of
the entanglement entropy of a single interval $[u_1,v_1]$, made of $n$ sheets
sequentially joined to each other on the segment $x\in[u_1,v_1],\,\tau=0$.
We expect that the associated partition function in a theory
defined on the complex plane $z=x+i\tau$
can be written in terms of certain ``fields'' at $z=v_1$ and $z=u_1$.

The partition function (here  ${\cal L}[\varphi](\rx,\ry)$ is the
local lagrangian {\it density}) \be \label{partfunct} Z_{\cal R} =
\int [d\varphi]_{{\cal R}} \exp\lt[-\int_{{\cal R}} d\rx
d\ry\,{\cal L}[\varphi](\rx,\ry)\rt]\,, \ee essentially defines
these fields, i.e. it gives their correlation functions, up to a
normalization independent of their positions. However in the model
on the complex plane, this definition makes them non-local (see
for a complete discussion \cite{ccd-07}). Locality is at the basis
of most of the results in field theory, so it is important to
recover it.

The solution to the problem consists in moving the complicated
topology of the world-sheet ${\cal R}$ (i.e. the space where the
coordinates $x,\ry$ live)  to the target space (i.e. the space
where the fields live). Let us consider a model formed by $n$
independent copies of the original model. (Note that $n$ is the
number of Riemann sheets necessary to describe the Riemann surface
by coordinates on the plane.) The partition function
(\ref{partfunct}) can be re-written as the path integral on the
complex plane \be \label{partfunctmulti} \fl Z_{\cal R} =
\int_{{\cal C}_{u_1,v_1}} \,[d\varphi_1 \cdots d\varphi_n]
\exp\lt[-\int_{\C}
d\rx d\ry\,({\cal L}[\varphi_1](\rx,\ry)+\ldots
+{\cal L}[\varphi_n](\rx,\ry))\rt]
\ee
where with $\int_{{\cal C}_{u_1,v_1}}$ we indicated the
{\it restricted}  path integral with conditions
\be
\varphi_i(\rx,0^+) = \varphi_{i+1}(\rx,0^-)~,\quad
\rx\in[u_1,v_1],\quad i=1,\ldots,n
\ee
where we identify $n+i\equiv i$. The lagrangian density of the
multi-copy model is
\[
{\cal L}^{(n)}[\varphi_1,\ldots,\varphi_n](\rx,\ry) =
{\cal L}[\varphi_1](\rx,\ry)+\ldots+{\cal L}[\varphi_n](\rx,\ry)
\]
so that the energy density is the sum of the energy
densities of the $n$ individual copies. Hence the expression
(\ref{partfunctmulti}) does indeed define local fields at
$(u_1,0)$ and $(v_1,0)$ in the multi-copy model \cite{ccd-07}.

The local fields defined in (\ref{partfunctmulti}) are examples of
``twist fields''.
Twist fields exist in a quantum field theory whenever there is a global
internal symmetry $\sigma$ (a symmetry that acts the same way everywhere in
space, and that does not change the positions of fields):
$\int d\rx d\ry\, {\cal L}[\sigma\varphi](\rx,\ry) =
\int d\rx d\ry\, {\cal L}[\varphi](\rx,\ry)$.
In the model with lagrangian ${\cal L}^{(n)}$, there is a symmetry under
exchange of the copies. The twist fields defined by (\ref{partfunctmulti}),
which have been called {\em branch-point twist fields} \cite{ccd-07}, are
twist fields associated to the two opposite cyclic permutation symmetries
$i\mapsto i+1$ and $i+1\mapsto i$  ($i=1,\ldots,n,\;n+1\equiv 1$).
We can denote them simply by $\tw_n$ and $\t\tw_n$, respectively
\bea
\tw_n\equiv\tw_\sigma~,\quad &\sigma&\;:\; i\mapsto i+1 \ {\rm mod} \,n\,, \\
\t\tw_n\equiv\tw_{\sigma^{-1}}~,\quad &\sigma^{-1}&\;:\;
i+1\mapsto i \ {\rm mod} \,n\,. \eea Notice that $\t\tw_n$ can be
identified with $\tw_{-n}$ (and in fact they were called $\Phi_n$
and $\Phi_{-n}$ in Ref.~\cite{cc-04}).

For the $n$-sheeted Riemann surface along the set $A$ made of
$N$ disjoint intervals $[u_j,v_j]$ we then have
\be
\label{pftwopt}
Z_{{\cal R}_{n,N}} \propto \bra
\tw_n(u_1,0) \t\tw_n(v_1,0)\cdots \tw_n(u_N,0) \t\tw_n(v_N,0) \ket_{{\cal L}^{(n)},\C}\,.
\ee
This can be seen by observing that for $\rx\in[u_j,v_j]$, consecutive copies
are connected through $\ry=0$ due to the presence of $\tw_n(v_j,0)$,
whereas for $\rx$ in $B$, copies are connected to themselves through
$\ry=0$ because the conditions arising from the definition of
$\tw_n(u_j,0)$ and $\t\tw_n(v_j,0)$ cancel each other.
More generally, the identification holds for correlation functions in the
model ${\cal L}$ on ${\cal R}_{n,1}$
\be\label{br-tw}
\fl \bra \Or(\rx,\ry; \mbox{sheet $i$}) \cdots \ket_{{\cal L},{\cal R}_{n,1}} =
  \frac{\bra \tw_n(u_1,0) \t\tw_n(v_1,0) \Or_i(\rx,\ry) \cdots \ket_{{\cal L}^{(n)},\C}}{\bra \tw_n(u_1,0) \t\tw_n(v_1,0) \ket_{{\cal L}^{(n)},\C}}
\ee
where $\Or_i$ is the field in the model ${\cal L}^{(n)}$ coming from
the $i^{\rm th}$ copy of ${\cal L}$, and the ratio properly takes into account
all the proportionality constants.
The same expression with the products of more twist and anti-twist fields
holds in the case of ${\cal R}_{N,n}$.

It is also useful to introduce the linear combination of the basic fields
\begin{equation}
\tilde{\varphi}_k \,\equiv\,
\sum_{j\,=\,1}^n e^{2\pi i \frac{k}{n} j} \varphi_{j}\,,\qquad
k\,=\,0,1, \dots , n-1\,,
\end{equation}
that get multiplied by $e^{2\pi ik/n}$ on going around the twist
operator, i.e. they diagonalize the twist \be \tw_n
\tilde{\varphi}_k=e^{2\pi ik/n} \tilde{\varphi}_k\,,\qquad {\rm
and}\qquad \t\tw_n \tilde{\varphi}_k=e^{-2\pi ik/n}
\tilde{\varphi}_k\,. \ee Notice that when the basic field
$\varphi_j$ are real $\tilde{\varphi}_k^* =
\tilde{\varphi}_{n-k}$. When the different values of $k$ decouple,
the total partition function is a product of the partition
functions for each $k$. Thus also the twist fields can be written
as products of fields acting only on  $\tilde{\varphi}_k$ \be
\tw_n=\prod_{k=0}^{n-1}\tw_{n,k}\,,\qquad
\t\tw_n=\prod_{k=0}^{n-1}\t\tw_{n,k}\,, \ee with $\tw_{k,n}
\tilde{\varphi}_{k'}= \tilde{\varphi}_k$ if $k\neq k'$ and
$\tw_{k,n} \tilde{\varphi}_{k}= e^{2\pi ik/n} \tilde{\varphi}_k$.
Thus \be Z_{\cal R}=\prod_{k=0}^{n-1} \bra
\tw_{k,n}(u_1,0)\t\tw_{k,n}(v_1,0) \dots\ket_{{\cal
L}^{(n)},\C}\,. \label{Tdiag} \ee This way of proceeding is useful
for free theories as in Refs.
\cite{ch-05,cfh-05,ch-07,chl-09,ch-rev}, when the various
$k$-modes decouple leading to Eq.~(\ref{Tdiag}).

\section{Entanglement entropy in conformal field theory: a single interval}
\label{sec3}

\begin{figure}
\includegraphics[width=0.7\textwidth]{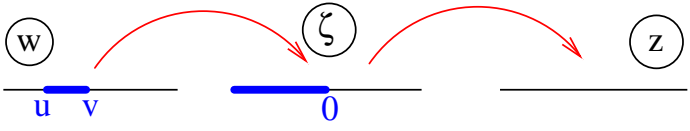}
\caption{Uniformizing transformation for ${\cal R}_{n,1}$.
$w\to\zeta=(w-u)/(w-v)$ maps the branch points to $(0,\infty)$.
This is uniformized by the mapping $\zeta\to z=\zeta^{1/n}$. }
\label{map}
\end{figure}

Following Ref.~\cite{cc-04}, we first consider the case $N=1$ and
no boundaries, that is the case considered by Holzhey et al.
\cite{Holzhey} of a single interval $[u,v]$ of length
$\ell\equiv|u-v|$ in an infinitely long 1D quantum system, at zero
temperature. The complex coordinate is $w=x+i\tau$ and  ${\bar
w}=x-i\tau$. The conformal mapping $w\to\zeta=(w-u)/(w-v)$ maps
the branch points to $(0,\infty)$. This is uniformized by the
mapping $\zeta\to z=\zeta^{1/n}= \big((w-u)/(w-v)\big)^{1/n}$.
This maps the whole of the $n$-sheeted Riemann surface ${\cal
R}_{n,1}$ to the $z$-plane ${\bf C}$, see Fig. ~\ref{map} for an
illustration of this. Now consider the holomorphic component of
the stress tensor $T(w)$. This is related to the transformed
stress tensor $T(z)$ by \cite{BPZ}
\begin{equation}
\label{schwartz}
T(w)=\left(\frac{dz}{dw}\right)^2 T(z)+\frac{c}{12}\{z,w\}\,,
\end{equation}
where $\{z,w\}=(z'''z'-\frac32{z''}^2)/{z'}^2$ is the Schwarzian
derivative. In particular, taking the expectation value of this,
and using $\langle T(z)\rangle_{\bf C}=0$ by translational and
rotational invariance, we find
\begin{equation}
\langle T(w)\rangle_{{\cal R}_{n,1}}=\frac c{12}\{z,w\}=
{c(1-n^{-2})\over 24}{(v-u)^2\over(w-u)^2(w-v)^2}\,.
\end{equation}
From Eq.~(\ref{br-tw}), this is equal to
$$
\frac{\bra\tw_n(u,0)\t\tw_n(v,0) T_j(w)\ket_{{\cal L}^{(n)},\C}}{
\bra \tw_n(u,0) \t\tw_n(v,0)\ket_{{\cal L}^{(n)},\C}}\,,
$$
for all $j$. We can then obtain the correlation function
involving the stress-energy tensor of ${\cal L}^{(n)}$ by multiplying by $n$:
\[
\frac{\bra \tw_n(u,0) \t\tw_n(v,0) T^{(n)}(w)\ket_{{\cal L}^{(n)},\C}}{\bra \tw_n(u,0) \tw_n(v,0)\ket_{{\cal L}^{(n)},\C}} =
    \frac{c(n^2-1)}{24 n} \frac{(u-v)^2}{(w-u)^2(w-v)^2}.
\]
The comparison with the conformal Ward identity \cite{BPZ}
\bea &&\fl
\bra \tw_n(u,0) \t\tw_n(v,0) T^{(n)}(w)\ket_{{\cal L}^{(n)},\C} = \nonumber\\ && \fl
\qquad
\left(\frac1{w-u} \frac{\partial}{\partial u} + \frac{h_{\tw_n}}{(w-u)^2} +
\frac1{w-v} \frac{\partial}{\partial v} + \frac{h_{\t\tw_n}}{(w-v)^2}\right)
\bra \tw_n(u,0) \t\tw_n(v,0)\ket_{{\cal L}^{(n)},\C}\,,
\eea
allows us to identify the scaling dimension of the primary fields
$\tw_n$ and $\t\tw_n$
(they have the same scaling dimension $d_n=\bar{d}_n$)
using $\bra \tw_n(u,0) \t\tw_n(v,0)\ket_{{\cal L}^{(n)},\C} = |u-v|^{-2d_n}$
\footnote{We use here $d_n$ instead of $\Delta_n$
in \cite{cc-04} to avoid
confusion between these scaling dimensions, in fact they are not the same and
are related by $d_n= 2n \Delta_n$.}
\be\label{scdim}
    d_n = \frac{c}{12} \left(n-\frac1n\right)~.
\ee To our knowledge this scaling dimension was first derived by
Knizhnik \cite{k-87} in a different context.

The above equation determines all the properties under conformal
transformations, and we therefore conclude that the renormalized
$Z_n({A})/Z^n\propto{\rm Tr}\,\rho_{A}^n$ behaves (apart from a
possible overall constant) under scale and conformal
transformations identically to the two-point function of a primary
operator with dimension $d_n$. In particular, this means that
\begin{equation}
{\rm Tr}\,\rho_{A}^n=c_n\left(\frac{v-u}{a}\right)^{- c(n-1/n)/6}\,.
\label{RenCFT}
\end{equation}
The power of $a$ (corresponding to the renormalization constant
$\cal Z$) has been inserted to make the final result
dimensionless, as it should be. The constants $c_n$  cannot be
determined with this method. However $c_1$ must be unity. The
analytic continuation is straightforward leading to the R\'enyi and
von Neumann entropies \be \fl \qquad
S_A^{(n)}=\frac{c}6\left(1+\frac1n\right) \log\frac{\ell}a
+c'_n\,, \qquad\qquad S_A=\frac{c}3 \log\frac{\ell}a +c'_1\,, \ee
where we defined the non universal constant \be
c'_n\equiv\frac{\log c_n}{1-n}\,. \ee 
Notice that $c'_1$ is minus the derivative of $c_n$ at $n=1$.
Despite their non-universal nature, the constants $c_n'$ are known exactly 
for few integrable models \cite{jk-04,ijk-05,fik-08,ij-08,ccd-07}.

Under certain conditions, the entanglement entropy can also be expressed
in terms of averages over ensembles of random matrices \cite{km-04}
providing a new connection
between the universality class of the conformal field theory and
random matrix ensembles.

Interestingly, Eq.~(\ref{RenCFT}) describes the asymptotic
behaviour for large enough $\ell$ for any $n>0$ (and not only for
$n\geq1$) at least for the simplest solvable models, where it can
be explicitly checked. This is of relevance for the convergence of
of the algorithms based on matrix product states \cite{mps,mps2}.

\subsection{Generalizations: Finite temperature or finite size}

The fact that ${\rm Tr}\,\rho_{A}^n$ transforms under a general
conformal transformation as a two-point function of primary operators
$\tw$ means that it is simple to compute in other geometries, obtained by a
conformal mapping $z\to w=w(z)$, using the formula
\be
\langle\tw_n(z_1,\bar z_1)\t\tw_n(z_2,\bar z_2)\rangle
=|w'(z_1)w'(z_2)|^{d_n}
\langle\tw_n(w_1,\bar w_1)\t\tw_n(w_2,\bar w_2)\rangle\,.
\label{2pttra}
\ee

Particularly relevant is the transformation $w\to z=(\beta/2\pi)\log w$ that
maps each sheet in the $w$-plane into an infinitely long cylinder of
circumference $\beta$.
The sheets are now sewn together along a branch cut joining the images
of the points $u$ and $v$. By arranging this to lie parallel to the axis
of the cylinder, we get an expression for $\Tr\,\rho_{A}^n$
in a thermal mixed state at finite temperature $\beta^{-1}$.
After simple algebra, this leads to the result for the
entropy \cite{cc-04,k-93}
\begin{equation}
S_A= \frac{c}3 \log\left(\frac{\beta}{\pi
a}\sinh\frac{\pi\ell}{\beta}\right)+c_1'= \cases{ \displaystyle
\frac{c}3 \log \frac{\ell}{a}+c_1' & $\ell\ll\beta$\;,\cr
\displaystyle  \frac{\pi c}{3\beta}\ell+c_1'    &
$\ell\gg\beta$\,. } \label{finiteT}
\end{equation}
This simple formula interpolates between the zero-temperature
result for $\ell\ll\beta$ and an extensive form in opposite limit
$\ell\gg\beta$. In this limit its density agrees with that of the
Gibbs entropy of an isolated system of length $\ell$, as obtained
from the standard CFT expression \cite{BCN,Affleck} $\beta F\sim
-(\pi c/6)(\ell/\beta)$ for its free energy. As expected, in the
high-temperature limit, the von Neumann entropy reduces to a pure
thermal form and the entanglement vanishes.

\begin{SCfigure}
\includegraphics[width=0.56\textwidth]{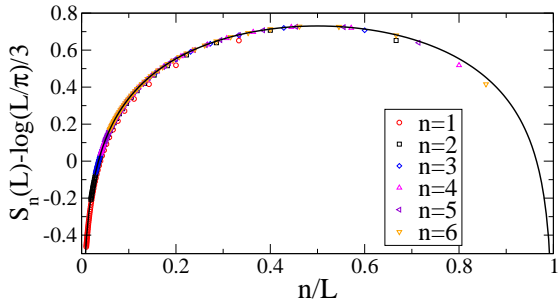}
\caption{Exact finite size scaling of the entanglement entropy $S_n(L)$ (here
  $n=\ell$) for the XXZ model at $\Delta=1/2$ against the
  CFT prediction $1/3 \log\sin (\pi n/L)+c'_1$ (full line).
  $c_1'=0.7305$ has been fixed \cite{ccn-08}.
  Reprinted with permission from \cite{ncc-08}.}
\label{SnL}
\end{SCfigure}

Alternatively, we may orient the branch cut perpendicular to the axis
of the cylinder, which, with the replacement $\beta\to L$, corresponds
to the entropy of a subsystem of length $\ell$ in a finite 1D system of
length $L$, with periodic boundary conditions, in its ground state.
This gives
\begin{equation}
\fl
{\rm Tr}\,\rho_{A}^n=
c_n\left(\frac{L}{\pi a}\sin\frac{\pi\ell}L\right)^{-c(n-1/n)/6}\,,
\quad \Rightarrow\quad
S_A= \frac{c}3\log\left(\frac{L}{\pi a}\sin\frac{\pi\ell}L\right)+c_1'\,.
\label{finiteL}
\end{equation}
$S_A$ is symmetric under $\ell\to L-\ell$ and it is maximal when
$\ell=L/2$. This relation is fundamental in the analysis of
numerical data that are mainly done in finite size. It provides an
unambiguous way to determine the central charge even from
relatively small system sizes. In Fig.~\ref{SnL} we report the
exact calculation of $S_A$ for the XXZ chain at $\Delta=1/2$ from
Ref.~\cite{ncc-08}, showing that already small values of $\ell$
($\leq6$) it gives the correct asymptotic scaling. We stress that the
most powerful aspect in the determination of the central charge
via the entanglement entropy is that it does not involve the a
priori knowledge of the speed of the sound, unlike other measures
based on the gap or free energy scaling. For this reason, this
method has been widely used in recent years.

\subsubsection{Finite temperature and finite size.}

When a finite system is also at finite temperature, we need to
consider periodic boundary conditions both on space and imaginary
time axes. This results in the calculation of a two-point function
of twist operators on a {\it torus}. As it is well known \cite{cftbook},
in this case correlations depend not only on the scaling
dimensions, but on the full operator content of the theory and the
calculations must be done for any universality class. Furthermore
it is not possible to use a uniformizing transformation from the
plane, because of the non-trivial topology of the torus.

To our knowledge, this calculation has been performed only for the
massless Dirac fermion \cite{ant-08}. In this case, it is
convenient to use the representation of the twist fields in the
diagonal basis (see Eq.~(\ref{Tdiag})) to obtain \be
\mbox{Tr}\rho_A^n=\prod_{k=0}^{n-1} \bra
\tw_{n,k}(z,\bar{z})\t\tw_{n,k}(0,0) \ket\,, \ee and these
two-point functions are known from bosonization (setting $z=\ell$
and $L=1$ as scale) \be \bra \tw_{n,k}(\ell)\t\tw_{n,k}(0)\ket=
\left|\frac{2\pi\eta(i\beta)^3}{\theta_1(\ell|i\beta)}\right|^{2k^2/n^2}
\frac{|\theta_\nu(k\ell/n|i\beta)|^2}{|\theta_\nu(0|i\beta)|^2},
\ee where $\theta_\nu$ and $\eta$ represent standard elliptic
functions, and $\nu=2,3,4$ is the sector of the fermion. This
gives a simple and compact answer for any integer $n$, but the
analytic continuation of the second part is complicated because
$k$ enters in the argument of the $\theta$ functions. For this
reason, it is possible to give exact expressions for the
entanglement entropy only in the high- and low-temperature
expansions (that however give  convergent expressions). As an
example we report here the high-temperature expansion in the NS
sector ($\nu=3$) from Ref.~\cite{ant-08} \bea \fl
S_{A}&=&\frac13\log\left[\frac{\beta}{\pi
a}\sinh\frac{\pi\ell}{\beta}\right]+\frac13\sum_{m=1}^\infty \log
\frac{(1-e^{2\pi \frac{\ell}{\beta}}e^{-2\pi \frac{m}{\beta}
})(1-e^{-2\pi \frac{\ell}{\beta}}e^{-2\pi
\frac{m}{\beta}})}{(1-e^{-2\pi \frac{m}{\beta}})^2}\nonumber \\
\fl&&\qquad + 2\sum_{l=1}^\infty
\frac{(-1)^{l}}{l}\cdot\frac{\frac{\pi \ell
l}{\beta}\coth\left(\frac{\pi \ell
l}{\beta}\right)-1}{\sinh\left(\pi \frac{l}{\beta}\right)}. \eea
This formula gives one example of the crossover form
Eq.~(\ref{finiteL}) for $\beta\gg1$ to Eq.~(\ref{finiteT}) for
$\beta\ll1$. More details about the derivation and the results for
other sectors can be found in Ref.~\cite{ant-08}.
We mention that these issues have been investigated numerically for 
different spin-chains in Ref. \cite{zpp-08}.

\subsection{Systems with boundaries}
In numerical simulations with DMRG and in real experimental life,
physical systems do not obey periodic boundary conditions, but rather have
some open boundaries. While in the study of correlation functions
of local operators the effect of the boundaries can be reduced by
performing measures far from them, the intrinsically global nature
of the block entanglement makes it more sensible to the boundary
conditions. This is not a negative feature and can be effectively
described by boundary CFT \cite{cardy-84,cardy-05}.

Let us start by considering a 1D system is a semi-infinite line,
say $[0,\infty)$, and the subsystem $A$ is the finite interval
$[0,\ell)$. The $n$-sheeted Riemann surface $\t{\cal R}_{n,1}$
then consists of $n$ copies of the half-plane $x\geq0$, sewn
together along $0\leq x<\ell, \tau=0$. Once again, we work
initially at zero temperature. It is convenient to use the complex
variable $w=\tau+ix$. The uniformizing transformation is now
$z=\big((w-i\ell)/(w+i\ell)\big)^{1/n}$, which maps the whole
Riemann surface to the unit disc $|z|\leq1$. In this geometry,
$\langle T(z)\rangle=0$ by rotational invariance, so that, using
(\ref{schwartz}), we find
\begin{equation}
\label{hp}
\langle T(w)\rangle_{\t{\cal R}_{n,1}}=\frac{c}{24}(1-n^{-2})
{(2\ell)^2\over(w-i\ell)^2(w+i\ell)^2}\,.
\end{equation}
Note that in the half-plane, $T$ and $\overline T$ are related by
analytic continuation: $\overline T(\bar w)=[T(w)]^*$
\cite{cardy-84}. Eq.~(\ref{hp}) has the same form as $\langle
T(w)\tw_n(i\ell)\rangle$, which follows from the Ward identities
of boundary CFT \cite{cardy-84}, with the normalization
$\langle\tw_n(i\ell)\rangle=(2\ell)^{-d_n}$.

The analysis then proceeds in analogy with the previous case leading to
\begin{equation}
{\rm Tr}\,\rho_{A}^n= \t c_n\left(\frac{2\ell}a\right)^{(c/12)(n-1/n)}\,
\Rightarrow\,
S_A=\frac{c}6\log\frac{2\ell}a+{\tilde c}'_1\,.
\label{Sbou}
\end{equation}
The constants $\t c_n$ are in principle different from $c_n$ in
the periodic case. The coefficient in front of the logarithm is
one-half of the one with periodic boundary conditions. This can be
interpreted as the analogue of the area law in 1D. In fact, while
with periodic conditions there are two boundary-points between $A$
and $B$, in the present case there is only one.

Once again, this result can be conformally transformed into a number of
other cases. At finite temperature $\beta^{-1}$ we find
\begin{equation}
S_A=\frac{c}6\log\left(\frac\beta{\pi a}\sinh\frac{2\pi\ell}\beta\right)+
{\tilde c}'_1=
\cases{
\displaystyle  \frac{c}6 \log \frac{2\ell}{a}+\t c_1' & $\ell\ll\beta$\;,\cr
\displaystyle  \frac{\pi c}{3\beta}\ell+\t c_1'    & $\ell\gg\beta$\,,
}
\end{equation}
In the limit $\ell\gg\beta$ we find the same extensive entropy as before.
This allows to identify \cite{cc-04,zbfs-06,lsca-06}
\be
{\t c}'_1-\frac{c'_1}2= \log g\,,
\label{Bent}
\ee
where $\log g$ is the boundary entropy, first discussed by
Affleck and Ludwig \cite{al-91}. $g$ depends only on the boundary CFT
and its value is known in the simplest cases.
Numerical simulations confirm with high precision this relation \cite{zbfs-06}.
It is worth mentioning that the change in the entanglement 
entropy of topological quantum Hall fluids (see the Fradkin review in this special issue
\cite{f-rev} for details) at a point of constriction is related to the change of Affleck and 
Ludwig entropy of the coupled edge states of the fluid at the point contact \cite{ffn-07}.

For a finite 1D system, of length $L$, at
zero temperature, divided into two pieces of lengths $\ell$ and
$L-\ell$, we similarly find
\begin{equation}
S_A=\frac{c}6\log\left(\frac{2L}{\pi a}\sin\frac{\pi\ell}L\right)+\t c'_1\,.
\end{equation}
This last equation is the most appropriate for numerical simulations that
are usually performed in finite systems with some boundary conditions
at both ends.

\subsubsection{Interfaces.}
We have seen that when a system is translationally invariant the entanglement
entropy scales like $S_A=c/3 \log\ell$, while in the presence of a boundary,
that can be a disconnected chain, it scales like $S_A=c/6 \log\ell$.
In a condensed matter system like a spin chain, we can modulate a single
bond (let say at $x=0$) from zero to the value in the rest of the chain,
going from one extreme to the other.
In the presence of such a {\it defect}, there are mainly three possibilities
under renormalization group:
\begin{itemize}
\item The defect is irrelevant: the system flows to the translational invariant
Hamiltonian and $S_A=c/3 \log \ell$.
\item The defect is relevant: the RG flow leads the system to a different fixed
point. In particular when non-trivial ones are excluded, it flows to a
disconnected system with $S_A=c/6 \log \ell$.
\item The defect is marginal: the critical properties, and in particular the
entanglement entropy, are continuous function of the defect strength.
\end{itemize}

It has been shown numerically \cite{def2} and analytically
\cite{l-04} that in the gapless phase of the XXZ chain with
$\Delta\neq0$, the defect is either relevant or irrelevant,
leading always to the well-known behaviors of $S_A$. More
interesting is the case of the XX model \cite{p-def}, when the
defect is marginal and for the entanglement entropy one gets \be
S_A= \frac{\sigma(t)}3 \log\ell\,, \ee where $t$ is the strength
of the defect ($t=0$ for disconnected chains and $t=1$ for
translational invariant ones), and $\sigma(t)$ is a monotonous
increasing function of $t$ with $\sigma(0)=1/2$ and $\sigma(1)=1$.
A similar behaviour has been found also for more complicated
defects in Ref.~\cite{isl-09}.

This phenomenon can be described as an interface between two
different CFT's with $c=1$. In Ref.~\cite{ss-08} the entanglement
entropy of two systems of length $L$ separated by an interface
with scattering amplitude $s$ (the analogous of $t$ above) has
been calculated \be\fl S_A=\sigma(|s|)\log L\,, \quad {\rm
with}\quad \sigma(|s|)= \frac{|s|}2-\frac2{\pi^2}\int_0^\infty
u(\sqrt{1+(|s|/\sinh u)^2}-1) du\,, \ee (the integral can also be
written in terms of polylog functions). When there is no
interface, i.e. for $s=1$, $\sigma(1)=1/3$, as expected. Instead
$\sigma(0)=0$ because the two CFT's decouple. Unfortunately no
result for a subsystem of length $\ell<L$ in the presence of the
interface is still available to be compared with the results in
the XX chain \cite{p-def}.

Other results for more general defects are known \cite{scla-07}, but
we remand to the review by Laflorencie et al. in this issue \cite{lsa-rev} for
an extensive discussion.

\subsection{General appearance of logarithmic behaviour}

In arbitrary dimension, the scale invariance (i.e. criticality of
the statistical model) together with translational and rotational
(i.e. Lorentz in real time) invariance and locality automatically
leads to conformal invariance \cite{cftbook}, explaining the very
large interest in these theories. However, nature is not always so
kind and there are physical systems that are critical, but that
because of the explicit breaking of translational and/or
rotational invariance are not conformal. It is then natural to ask
what is the behaviour of the entanglement entropy in these
systems. Srednicki \cite{s-93} argued that the area law in higher
dimensional systems for a gapless model should generally collapse
to a $\log\ell$ behaviour in one dimension, and so one would
expect the appearance of logarithms even in non-conformal
invariant systems. Unfortunately, nowadays there are several
examples of the breaking of the area scaling in critical systems
(see e.g. Ref.~\cite{e-rev}), leaving doubts on the earlier argument.

Critical systems showing the breaking of translational or rotational invariance
have been largely studied. We give here few important examples.
When translational invariance is broken by quenched disorder, the resulting
statistical model can be studied by strong-disorder RG methods and numerically.
In all the studied models, it has been shown unambiguously that the
entanglement entropy always shows a $\log\ell$ behaviour
\cite{rm-04,l-05,dmcf-06,s-06,by-07,bdmr-07,rm-07,il-08,frbm-08}
(see the review by Moore and Refael \cite{mr-rev} in this issue).
Translational invariance can also be broken by taking aperiodic couplings:
even in this case a $\log\ell$ behaviour has been found \cite{ijz-07,jz-07}.
Non-relativistic dispersion relations like $E=k^2$ also naturally breaks
conformal invariance, by breaking Lorentz. A well-known and physical important
example is the ferromagnetic Heisenberg chain, for which the entanglement
entropy scales like $S_A=1/2 \log\ell$ \cite{ps-05,pss-05}.
Another interesting example of this kind can be found in \cite{gh-09}.

Often it has been proposed that the scaling of the entanglement
entropy as $\log\ell$ can be used to define an effective central
charge for non conformally invariant systems that can share some
of the properties of the real central charge (as for example the
monotonicity along renormalization group flow \cite{Zam}). However,
conformal invariance is so powerful that fixes the scaling 
of the entanglement entropy, but, as we have seen, it also gives precise 
predictions for finite size scaling in Eq.~(\ref{finiteL}) and for the scaling 
of the R\'enyi entropies (\ref{RenCFT}). 
Before arguing about asymptotic restoration of
conformal invariance, all these relations should be carefully
checked. For example the finite size scaling found in
Ref.~\cite{pss-05,gh-09} is different from Eq.~(\ref{finiteL}).
Also the scaling of the entanglement entropy in the zero
temperature mixed state of the XXZ chain at $\Delta=1/2$ scales
logarithmically, but has a finite size form different from
Eq.~(\ref{finiteL}) \cite{ncc-08} and cannot be described by CFT.
Some excited states in spin-chains also displays logarithmic behavior because
 they can be interpreted as ground-states of properly defined
conformal Hamiltonian \cite{afc-09}.   
For random systems with quenched disorder, the finite size scaling form 
seems to be conformal from numerical simulations, but the R\'enyi entropies 
have different scaling in the random singlet phase \cite{mr-rev}.
We also mention that in several collective models a similar logarithmic 
behavior has been found \cite{coll}, but its origin is different from 
the one discussed here because of the absence of a spatial structure in these
mean-field like models. Their properties are related to those of the 
particle partitioning reviewed by Haque et al. in this volume \cite{hzs-rev}.

\section{Entanglement of disjoint intervals}
\label{sec4}

When the subsystem $A$ consists of several disjoint intervals, the
analysis becomes more complicated. In Ref.~\cite{cc-04} we
provided a result that in general is incorrect. This was based on
a uniformizing transformation mapping ${\cal R}_{n,N}$ into the
complex plane. However, the surface ${\cal R}_{n,N}$ has genus
$(n-1)(N-1)$ and so for $N\neq1$ (that is the case we already
discussed) cannot be uniformized to the complex plane (at the
level of the transformation itself, this has been discussed in
details \cite{cg-08}). The case $n=N=2$ has the topology of a
torus, whose partition function depends on the whole operator
content of the theory and not only on the central charge.
Consequently the simple formulas of Ref.~\cite{cc-04} cannot be
generally correct. The partition functions on Riemann surfaces
with higher genus are even more complicated.

Let us then start our analysis from the case of two intervals given
by the surface ${\cal R}_{n,2}$. By global conformal invariance the
partition function (that is the four-point correlation of twist fields)
can be written in the form (in this section we adsorb the normalization
$Z^n$ into $Z_{\mathcal{R}_{n,N}}$)
\be\fl
\Tr \rho_A^n\equiv Z_{\mathcal{R}_{n,2}}=c_n^2
\left(\frac{|u_1-u_2||v_1-v_2|}{|u_1-v_1||u_2-v_2||u_1-v_2||u_2-v_1|}
\right)^{2d_n} {\cal F}_{n}(x)
\label{Fn}
\ee
where $x$ is the real four-point ratio
\begin{equation}
x\,\equiv\,\frac{z_{12} \,z_{34}}{z_{13} \, z_{24}}=
\frac{(u_1-v_1)(u_2-v_2)}{(u_1-u_2)(v_1-v_2)}\,,
\end{equation}
and $d_n$ is given by Eq.~(\ref{scdim}). It can also be written as
\be Z_{\mathcal{R}_{n,2}}= Z_{\mathcal{R}_{n,2}}^W {\cal
F}_{n}(x)\,, \ee where $Z_{\mathcal{R}_{n,2}}^W$ is the incorrect
result in \cite{cc-04}. We normalized such that ${\cal
F}_{n}(0)=1$ (for $x\to0$, $Z_{\mathcal{R}_{n,2}}$ is the product
of the two two-point functions previously calculated and
normalized with $c_n$). The function ${\cal F}_{n}(x)$ depends
explicitly on the full operator content of the theory and must be
calculated case by case.

In Ref.~\cite{fps-08}, using old results of CFT on orbifolded
space \cite{Dixon,z-87}, ${\cal F}_{2}(x)$ has been calculated for
the Luttinger liquid CFT, that is a free boson compactified on a 
circle of radius $R$ 
\be {\cal
F}_{2}(x)= \frac{\theta_3 (\eta \tau) \theta_3 (\tau/\eta)}{
[\theta_3 (\tau)]^{2}}, \label{F2} 
\ee 
where $\tau$ is pure-imaginary, and is related to $x$ via 
$x= [\theta_2(\tau)/\theta_3(\tau)]^4$. $\theta_\nu$ are Jacobi theta
functions. $\eta$ is proportional to the square of the
compactification radius $R$ (while the definition of $R$ usually
depends on the normalization of the action, $\eta$ is the same in
all literature, that is why we prefer to write everything only in
terms of $\eta$ that usually is written as $\eta=1/(2K)$ in Luttinger liquid
notation).

\begin{figure}[t]
\includegraphics[width=0.445\textwidth]{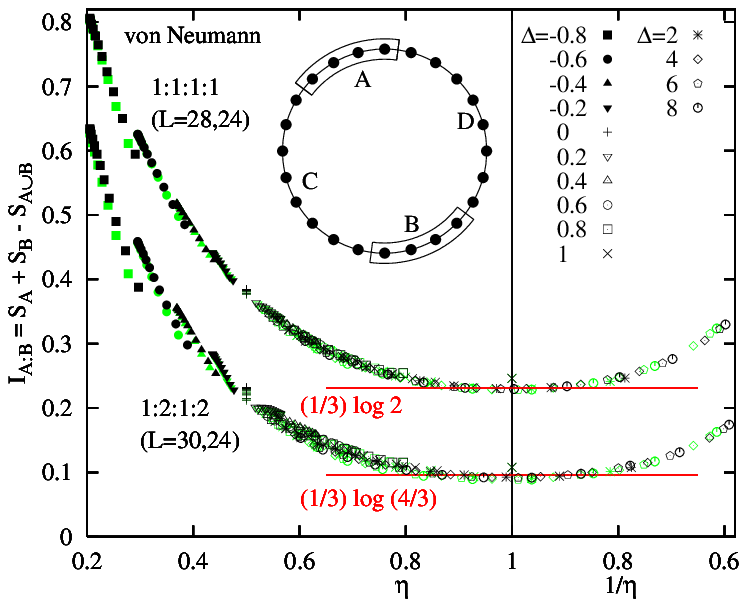}
\hspace{0.02\textwidth}
\includegraphics[width=0.52\textwidth]{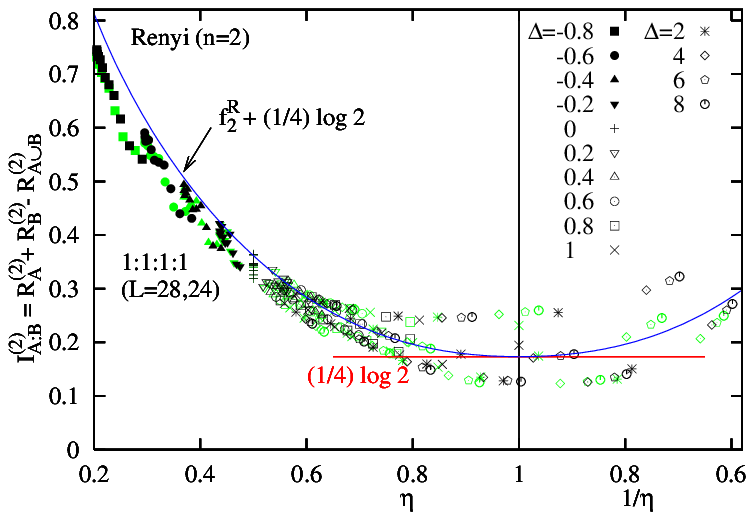}
\caption{The mutual information for fixed four-point ratio $x$ as
function of $\eta$ in the gapless phase of the XXZ model. The
horizontal lines stand for $Z_{\mathcal{R}_{n,2}}^W$ of
Ref.~\cite{cc-04}. Left: mutual information of the von Neumann
entropy. Right:  mutual information of the R\'enyi entropy for
$n=2$, compared with the compactified boson prediction
Eq.~(\ref{F2}). Reprinted with permission from \cite{fps-08}}
\label{Figmul}
\end{figure}

In order to check this prediction and the failure of
$Z_{\mathcal{R}_{n,2}}^W$, in Ref.~\cite{fps-08} the entanglement
of the XXZ chain for generic values of the anisotropy $\Delta$ and
magnetic field always in the gapless phase has been calculated by
direct diagonalization for systems up to 30 spins. In the absence
of the magnetic field, $\eta$ is related to the anisotropy by
$\eta=1-(\arccos\Delta)/\pi$, while for non-zero $h_z$ a closed
formula for $\eta$ does not exist and must be calculated
numerically as explained in \cite{fps-08}. The main results coming
from the exact diagonalization are reported in Fig.~\ref{Figmul}
in terms of the R\'enyi mutual information \be I_{A_1:A_2}^{(n)}=
S_{A_1}^{(n)}+S_{A_2}^{(n)}-S_{A_1\cup A_2}^{(n)}\,, \ee where
$A_1$ and $A_2$ are the two intervals composing $A=A_1\cup A_2$.
In Fig.~\ref{Figmul} the mutual information for $n=1,2$ is
reported. The $\eta$ dependence in both the cases is evident, and
the good collapse of the data shows the correctness of the scaling
in Eq.~(\ref{Fn}). (In $I_{A_1:A_2}^{(2)}$ the collapse is worst
due to the oscillating corrections to the scaling of the R\'enyi
entropies, known already for the single interval \cite{ccn-08}).
In the right panel of Fig.~\ref{Figmul}, the comparison of the
numerical results with the prediction (\ref{F2}) is reported. The
agreement is satisfactory, considering the small subsystem sizes
and the strong oscillations. The results for $I_{A_1:A_2}^{(1)}$
in the left panel of  Fig. ~\ref{Figmul} are more stable because
of the absence of oscillations. These are then more suitable for
the comparison with an analytic calculation.

In Ref.~\cite{cct-09} (in collaboration with E.~Tonni), we managed
to calculate ${\cal F}_{n}(x)$ for generic integral $n\geq1$. The
result reads
\begin{equation}
\mathcal{F}_n(x)=
\frac{\Theta\big(0|\eta\Gamma\big)\,\Theta\big(0|\Gamma/\eta\big)}{
\Theta\big(0|\Gamma\big)^2}\,,
\label{Fnv}
\end{equation}
where $\Gamma$ is an $(n-1)\times(n-1)$ matrix with elements
\be
\Gamma_{rs} =
\frac{2i}{n} \sum_{k\,=\,1}^{n-1}
\sin\left(\pi\frac{k}{n}\right)\beta_{k/n}\cos\left[2\pi\frac{k}{n}(r-s)\right],
\ee
and
\be
\beta_y=\frac{F_{y}(1-x)}{F_{y}(x)}\,,\qquad
F_{y}(x)\,\equiv\, _2 F_1(y,1-y;1;x)\,.
\ee
$\eta$ is exactly the same as above, while $\Theta$ is the Riemann-Siegel
theta function
\begin{equation}
\label{theta Riemann def}
\Theta(z|\Gamma)\,\equiv\,
\sum_{m \,\in\,\mathbf{Z}^{n-1}}
\exp\big[\,i\pi\,m\cdot \Gamma \cdot m+2\pi i \,m\cdot z\,\big]\,,
\end{equation}
that for $n-1=1$ reduces to the Jacobi $\theta_3(\tau=i\beta_{1/2})$, 
reproducing Eq.~(\ref{F2}).

Unfortunately we have been not yet able to continue analytically
this result to real $n$ for generic $\eta$ and so obtain the
entanglement entropy to compare with the left panel of
Fig.~\ref{Figmul} from Ref.~\cite{fps-08}.

Some interesting properties can be readily deduced from
Eq.~(\ref{Fnv}):
\begin{itemize}
\item For any $n$, it is symmetric under the exchange
$\eta\leftrightarrow 1/\eta$, generalizing the result for $n=2$ in
Eq.~(\ref{F2}).
\item For $\eta=1$ the result $Z_{\mathcal{R}_{n,2}}^W$ is correct,
i.e. $\mathcal{F}_n(x)=1$.
\item It is symmetric under the exchange $x\leftrightarrow 1-x$ (crossing 
symmetry).
This the scaling limit of $S_A=S_B$ for finite systems~\cite{fps-08}.
\end{itemize}
These three important properties carry over to the analytic continuation
and so must be true for the Von-Neumann entropy at $n=1$.
These findings then explain the numerical results of Ref.~\cite{fps-08} (in the 
left panel of Fig.~\ref{Figmul} the symmetry $\eta\leftrightarrow 1/\eta$
and $\mathcal{F}'_1(x)=1$ are evident).

More details and other interesting properties can be found in
Ref.~\cite{cct-09}. We  discuss here the so-called uncompactified
limit for $\eta\gg1$ or by symmetry $\eta\ll1$. In this case we
have \cite{cct-09} for $\eta\ll1$
\begin{equation}\fl
{\cal F}_n(x)= \frac{\eta^{-(n-1)/2}}{
\big[\prod_{k=1}^{n-1}F_{k/n}(x)F_{k/n}(1-x)\big]^{1/2}}\equiv
\frac{\eta^{-(n-1)/2}}{\big[D_{n}(x)D_{n}(1-x)]^{1/2}}\,,
\label{decomp} \ee where the function
$D_n(x)=\prod_{k=1}^{n-1}F_{k/n}(x)$ has been analytically
continued in Ref.~\cite{cct-09}. Then the prediction \be
I_{A_1:A_2}^{(1)}(\eta\ll1)- I_{A_1:A_2}^{(1),W}\simeq -\frac12
\ln \eta +\frac{D_1'(x)+D_1'(1-x)}2\,, \label{Idec} \ee perfectly
agrees with the numerical results in Fig.~\ref{Figmul} (from
Ref.~\cite{fps-08}) for $\eta\leq0.4$ (again in the previous
equation $I_{A_1:A_2}^{(1),W}$ is the result of
Ref.~\cite{cc-04}). 
Also the regime $x\ll 1$ can be studied analytically \cite{cct-09}.

A few comments are now in order. In
Ref.~\cite{cfh-05,ch-08,ch-09}, the entanglement entropy for two
disjoint intervals has been calculated for a free the Dirac
fermion, that corresponds to a compactified boson with $\eta=1/2$
\cite{cftbook}. However, it has been found that the entanglement
entropy is given by $Z_{\mathcal{R}_{n,2}}^W$, in contrast with
the numerical calculations in Ref.~\cite{fps-08} and the analytic
one in \cite{cct-09}. The details of this apparent disagreement are still
not completely understood, but they should be traced back to the different boundary conditions
that result from constructing the reduced density matrix for spin or fermion variables. 
Another calculation in agreement with $Z_{\mathcal{R}_{n,2}}^W$ can 
be found in \cite{ch-04}. For the Ising model numerical computations \cite{ffip-08} 
also show a good agreement with $Z_{\mathcal{R}_{n,2}}^W$. Also in this case, it is
likely that the deviations from $Z_{\mathcal{R}_{n,2}}^W$ should be attributed to the 
choice of the variables used in constructing the reduced density matrix. (In fact,  
calculations in the spin variables \cite{atct-p} show numerically and analytically 
that $Z_{\mathcal{R}_{n,2}}^W$  is not correct.) 
Finally, holographic calculations in AdS/CFT correspondence \cite{rt-06,vr-08}, 
considering the classical limit in the gravity sector, also found 
$Z^W_{R_{n,2}}$. It would be interesting to understand how the correct 
result might arise from taking into account the quantum effects on the 
gravity side (for details see the review of Nishioka, Ryu, and
Takayanagi in this volume \cite{rt-rev}).

For general $N>2$, there are still no firm results in the literature.
By global conformal invariance one can deduce \be \label{generaln}
{\rm Tr}\,\rho_{A}^n=
c_n^N\left({\prod_{j<k}(u_k-u_j)(v_k-v_j)\over\prod_{j,
k}(v_k-u_j)} \right)^{(c/6)(n-1/n)}{\cal F}_{n,N}(\{ x\}) \,. \ee
For ${\cal F}_{n,N}(\{ x\})=1$ this is the incorrect result of
Ref.~\cite{cc-04} (note a typo in the denominator). $\{ x\}$
stands for the collection of $2N-3$ independent ratios that can be
built with $2N$ points. Some old results from CFT on orbifold in
Refs. \cite{z-87,br-87} could be useful to calculate ${\cal
F}_{n,N}(\{ x\})$ for a compactified boson.
We also mention that  in 2D systems with conformal invariant wave-function 
(reviewed in \cite{f-rev} in this volume), the entanglement
entropy of a single region displays an addictive universal term depending 
on $\eta$ \cite{2d}.

Finally, it is worth to recall that in the case of more intervals, the
entanglement entropy measures only the entanglement of the
intervals with the rest of the system.
It is {\it not} a measure of the entanglement of one interval with respect
to the others, that instead requires the introduction of more complicated
quantities because $A_1\cup A_2$ is in a mixed state (see e.g. Refs.
\cite{Neg} for a discussion of this and examples).

\section{Entanglement entropy in non-critical 1+1-dimensional models}
\label{sec5}

When a one-dimensional statistical model has a gap (i.e. the
underlying quantum field theory is massive) the entanglement
entropy saturates to a finite value \cite{Vidal}. This is an
analogue of the area law in one dimension, because the area is
only a number that does not increase with subsystem size, in
contrast to higher dimensions. Generally this value is very
complicated to be calculated and it is
known only in very simple cases. 
However, when a system is close to a {\it conformal} quantum critical point,
that is when the gap $\Delta$ is small (or the correlation length $\xi\propto
\Delta^{-1}$ is large) it is possible to derive a very general scaling
form \cite{cc-04}, that can be used also as an operative definition of the
correlation length.
Hastings \cite{h-07} (see also \cite{h-07a}) provided a rigorous proof of
the area-law for one-dimensional systems with a generic local Hamiltonian,
not necessarily close to a conformal critical point.

We consider an infinite non-critical
model in 1+1-dimensions, in the scaling limit where the lattice spacing
$a\to0$ with the correlation length (inverse mass) fixed.
This corresponds to a massive relativistic QFT. We first consider the
case when the subset $A$ is the negative real axis, so that the
appropriate Riemann surface has branch points of order $n$ at 0 and
infinity. However, for the non-critical case, the branch point at
infinity is unimportant: we should arrive at the same expression
by considering a finite system whose length $L$ is much greater than
$\xi$.

Our argument parallels that of Zamolodchikov \cite{Zam} for the proof of his
famous $c$-theorem.
Let us consider the expectation value of the stress tensor $T_{\mu\nu}$
of a massive euclidean QFT on such a Riemann surface. In complex
coordinates, there are three non-zero components: $T\equiv T_{zz}$,
$\overline T\equiv T_{\bar z\bar z}$, and the trace
$\Theta=4T_{z\bar z}=4T_{\bar zz}$. These are related by the
conservation equations
\be
\label{cons}
\partial_{\bar z}T+\frac14\partial_z\Theta=0\,
\quad {\rm and}\quad
\partial_z\overline T+\frac14\partial_{\bar z}\Theta=0\,.
\ee
Consider the expectation values of these components. In the
single-sheeted geometry, $\langle T\rangle$ and $\langle\overline
T\rangle$ both vanish, but $\langle\Theta\rangle$ is constant and
non-vanishing: it measures the explicit breaking of scale invariance in
the non-critical system. In the $n$-sheeted geometry, however, they
all acquire a non-trivial spatial dependence. By
rotational invariance about the origin, they have the form
\begin{eqnarray}
\langle T(z,\bar z)\rangle&=& F_n(z\bar z)/z^2\,,\\
\langle\Theta(z,\bar z)\rangle-\langle\Theta\rangle_1&=&
G_n(z\bar z)/(z\bar z)\,,\\
\langle\overline T(z,\bar z)\rangle&=& F_n(z\bar z)/{\bar z}^2 \,.
\end{eqnarray}
From the conservation conditions (\ref{cons}) we have
\begin{equation}
(z\bar z)\left(F'_n+\frac14G'_n\right)=\frac14G_n\,.
\end{equation}
Now we expect that $F_n$ and $G_n$ both approach zero exponentially fast for
$|z|\gg\xi$, while in the opposite limit, on distance scales $\ll\xi$,
they approach the CFT values
$F_n\to (c/24)(1-n^{-2})$, $G_n\to0$.

We define an effective $C$-function
\begin{equation}
\fl C_n(R^2)\equiv \left(F_n(R^2)+\frac14G_n(R^2)\right)\,
\quad\Rightarrow\quad
R^2{\partial\over\partial(R^2)}C_n(R^2)=\frac14G_n(R^2)\,.
\end{equation}
whose integral,
assuming that theory is trivial in the infrared (if the RG flow is
towards a non-trivial theory, $c$ should be replaced by $c_{UV}-c_{IR}$),
gives
\begin{equation}
\int_0^\infty {G_n(R^2)\over R^2}d(R^2)=-\frac{c}6\left(1-\frac1{n^{2}}\right)\,,
\end{equation}
or equivalently
\begin{equation}
\int\left(\langle\Theta\rangle_n-\langle\Theta\rangle_1\right)d^2\!R
=-\pi n \frac{c}6\left(1-\frac1{n^{2}}\right)\,,
\end{equation}
where the integral is over the whole of the $n$-sheeted
surface. Now this integral (multiplied by a factor $1/2\pi$
corresponding to the conventional normalization of the stress
tensor) measures the response of the free energy $-\log Z$ to a
scale transformation, i.e. to a change in the mass $m$, since this
is the only dimensionful parameter of the renormalized theory.
Thus the left hand side is equal to
\begin{equation}
-(2\pi)\,m\frac{\partial}{\partial m}\left[\log Z_n-n\log Z\right]\,,
\end{equation}
giving finally
\begin{equation}
{Z_n\over Z^n}= c_n(ma)^{(c/12)(n-1/n)}\,,
\label{Zmass}
\end{equation}
where $c_n$ is a constant (with however $c_1=1$), and we have
inserted a power of $a$, corresponding to the renormalization
constant $\cal Z$ discussed earlier, to make the result
dimensionless.
Differentiating at $n=1$, we find
\begin{equation}
\label{mass}
S_{A}=-\frac{c}6\log(ma)=\frac{c}6\log\frac{\xi}{a}\,,
\end{equation}
where $\xi$ is the correlation length. We re-emphasize that this
result was obtained only for the scaling limit $\xi\gg a$.

So far we have considered the simplest geometry in the which set $A$ and
its complement $B$ are semi-infinite intervals. The more general case,
when $A$ is a union of disjoint intervals, is more difficult in the
massive case. However it is still true that the entropy can be expressed
in terms of the derivative at $n=1$ of correlators of twist operators
$\tw,\t\tw$. The above calculation can be thought of in terms of the
one-point function $\langle\tw_n\rangle$. In any quantum field theory
a more general correlator
$\langle\prod_{i=1}^{k}\Phi (w_i)\rangle$, with $\Phi=\tw_n$ or $\t\tw_n$,
should obey cluster
decomposition: that is, for separations $|w_i-w_j|$ all $\gg\xi$, it
should approach $\langle\tw_n\rangle^{k}$. This suggests that, in this
limit, the entropy should behave as
\be
S_A={\cal A}\frac{c}{6}\log\frac{\xi}a\,,
\ee
where ${\cal A}=k$ is the number of boundary points between $A$ and its
complement. This would be the 1D version of the area
law. When the interval lengths are of the order of
$\xi$, we expect to see a complicated but universal scaling form for the
cross-over.

This scaling has been confirmed in several cases with ${\cal A}=1$
or $2$ (see e.g.
\cite{cc-04,p-04,ijk-05,w-06,fijk-07,fik-08,imm-08,eer-09,gl-rev}, but this
list is far from being exhaustive). The corrections to this
formula for $\ell\ll \xi$ are also universal
\cite{ccd-07,cd-08,d-08,cd-09} and are discussed in details in the
review by Castro-Alvaredo and Doyon \cite{cd-rev} in this issue.

\section{Entanglement spectrum}
\label{sec6}

The knowledge of the scaling form for $\Tr \rho_A^n$ as in
Eqs.~(\ref{RenCFT},\ref{finiteL},\ref{Sbou},\ref{Zmass}) gives
more information about the reduced density matrix than the
entanglement entropy. We have seen that in many cases it scales
like \be R_n\equiv\Tr \rho_A^n=c_n e^{-b(n-1/n)}\,, \label{Ral}
\ee with $b>0$ only depending on the main features of the set $A$,
on the characteristic length of the system $L_{\rm eff}$ (i.e.
$\ell,\xi, L\sin \pi \ell/L\dots$) and on the central charge. This
suggests that many properties of the reduced density matrix are
very universal and do not depend on the details of the theory. For
example, the scaling of the largest eigenvalues $ \l_{\rm max}$ of
$\rho_A$ is obtained by taking the limit for $n\to \infty$
\cite{pz-05}: $S_A^{(\infty)}=-\ln \l_{\rm max}=S_A/2$ defines the
so-called ``single-copy entanglement'' \cite{ec-05} and gives
another measure of the entanglement content of an extended system.
This peculiar scaling led to the conclusion that
``half the entanglement in critical systems is distillable from
a single specimen'' \cite{olec-06}.

This result is however only the tip of a lot of information about
the reduced density matrix obtainable from Eq.~(\ref{Ral}). This
information is encoded in the full spectrum of the reduced density
matrix, which has been shortly called ``entanglement spectrum''
\cite{lh-08} and has been derived in Ref.~\cite{cl-08} for 1D
systems from Eq.~(\ref{Ral}).

In order to characterize the entanglement spectrum, let us define the
distribution of eigenvalues $P(\l)=\sum_i \delta(\l-\l_i)$.
If we ignore the $n$ dependence of the coefficient $c_n$ (that however is
expected only to give corrections the leading behaviour), it is easy to
compute the entanglement spectrum by inverse Laplace transforming $R_n$,
obtaining \cite{cl-08}
\be
P(\l)=\delta(\l_{\rm max}-\l)+
\frac{b\,\theta(\lm-\l)}{\l\sqrt{b\ln(\l_{\rm max}/\l)}}
I_1(2\sqrt{b \ln(\l_{\rm max}/\l)})\label{PL}\,,
\ee
where $I_k(x)$ stands for the modified Bessel function of the first kind.
Amazingly, $P(\l)$ depends only on $\l_{\rm max}$ (we
recall $b=-\ln \l_{\rm max}$) and not on any other detail of the theory.
For some integrable gapped systems $P(\l\ll1)$ was previously
derived \cite{jap}.

Let us discuss now the main properties of $P(\l)$:
\begin{itemize}
\item{\it The mean number of eigenvalues} larger than a given $\l$ is
\be
n(\l)=\int_\l^{\l_{\rm max}} d\l P(\l) =
I_0(2\sqrt{b \ln (\lm/\l)})\,.
\label{nlam}
\ee
Note that for $\l\to 0$, $n(\l)$ diverges, as it should,  because in the
continuum the number of eigenvalues is infinite. In the lattice models,
this can be regularized by the finite number of degrees of freedom.
\item
{\it The normalization} $\sum \l_i=1$ corresponds to $\int \l P(\l) d\l=1$.

\begin{figure}
\includegraphics[width=.97\textwidth, clip]{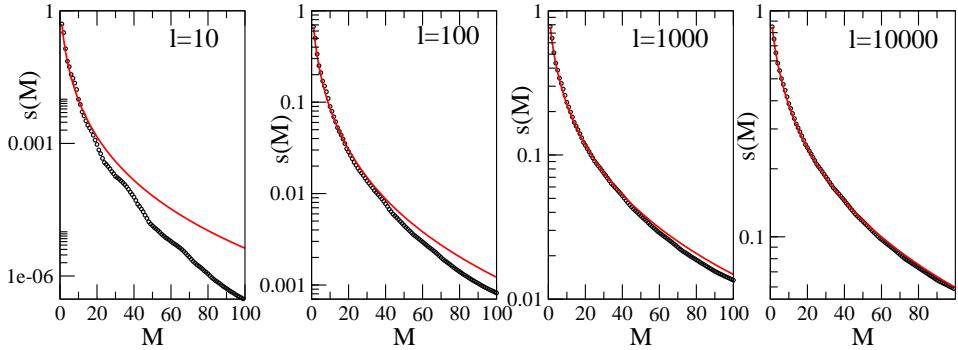}
\caption{Sum of the first $M$ eigenvalues of the XX model up to
$M=100$: $1-s(M)$ as function of $M$ for $\ell=10,100,1000,10000$
(black dots). The red line is the conformal field theory
prediction Eq.~(\ref{srformula}), in which $\lm$ has been fixed to
the maximum eigenvalue obtained numerically. Numerical data from
Ref.~\cite{cl-08}.} \label{sumrule}
\end{figure}
\item
{\it The entanglement entropy} is given by
\be
S= -\int_0^\lm \l \ln\l P(\l) d\l=-2\ln \lm\,, 
\ee 
reproducing the result that the single copy entanglement equals one-half of the
entanglement entropy.

\item
{\it Majorization} is a relation between two probability
distributions $\l\equiv\{\l_i\}$ and $\mu\equiv\{\mu_i\}$ whose
elements are ordered $\l_1>\l_2\dots >\l_N$ (and similarly for
$\mu$): it is said that $\l$ majorizes $\mu$ if $\sum_{i=1}^M \l_i
\geq\sum_{i=1}^M \mu_i $ for any $M=1,\dots,N$ and $\sum_{i=1}^N
\l_i =\sum_{i=1}^N \mu_i=1$. It has been argued, observed
numerically and in some instances proven analytically, that with
increasing $L_{\rm eff}$ the resulting distribution of eigenvalues
is majorized by the ones at smaller scaling lengths
\cite{maj,zbfs-06}.
Majorization follows easily from Eq.~(\ref{PL}): \be s(M)\equiv
\sum_{i=1}^M \l_i \to \lm\left[1+\int_0^{I_0^{-1}(M)} dy
e^{-y^2/4b} I_1(y)\right], \label{srformula} \ee at fixed $M$, is
a monotonous function of $\lm$ (that is a monotonous function of
$L_{\rm eff}$). This proves majorization.
\item The ratio of the first two eigenvalues (see \cite{cl-08} for details) is
\be
\frac{\l_2}{\l_{\rm max}}=e^{k/b}=e^{-\frac{6k}{c\ln(\ell/a)}}\,,
\ee
where $k$ is a constant and in the second equality we used the result for
periodic boundary conditions $b=-\ln\l_{\rm max}=(c/6)\ln (\ell/a)$.
This agrees with an old result for the scaling of eigenvalues of the
corner transfer matrix \cite{pt-87}.
\end{itemize}

Fig.~\ref{sumrule} shows an explicit check of the distribution of
eigenvalues for the XX model obtained in Ref.~\cite{cl-08} by
using the methods of Refs. \cite{p-dm,pe-rev}. It is evident that
when the subsystem size $\ell$ is large enough, the numerical
results perfectly agree with Eq.~(\ref{srformula}).

The knowledge of the scaling form of the entanglement spectrum has
been fundamental in understanding the convergence and the scaling
of the algorithms based on matrix product states (MPS) \cite{mps},
like DMRG. In these algorithms, the maximum amount of entanglement
that can be effectively described is limited by the dimension
$\chi$ of the matrix used to describe the state. The maximum
possible entanglement of this state is $S_{\rm max}= \log \chi$,
when all the components have the same weight $1/\chi$. But this
maximum-entanglement state has nothing to do with the ground-state
of the local Hamiltonian the algorithm is searching for, because
it is described by Eq.~(\ref{PL}). Numerical studies \cite
{t-08,pmtm-08}, in fact, showed that the entanglement of the MPS
approximating the critical ground-state scales like \be
S_A=\frac{c\kappa}6 \log \chi\,, \label{chiscaling} \ee defining
an effective length $\xi_\chi\sim \chi^{\kappa}$ \cite{t-08}.
$\kappa$ has been introduced as a new critical exponent of the MPS
\cite{t-08}. Using Eq.~(\ref{nlam}), it has been possible to
calculate this exponent exactly \cite{pmtm-08}, obtaining \be
\kappa=\frac{6}{c\sqrt{12/c+1}}\,, \ee and to show that the
corrections to Eq.~(\ref{chiscaling}) scale like $1/\log\chi$.

It is worth mentioning that a new numerical algorithm specifically
based on the scaling properties of the entanglement in a conformal
critical point has been recently proposed by Vidal \cite{v-07}:
the multi-scale entanglement renormalization ansatz (MERA). In a
MERA, the ground-state of an extended quantum system is organized
in layers corresponding to different length scales and, at a
quantum critical point, each layer equally contributes to the
entanglement of a block. This method then allowed to simulate
systems of remarkably large sizes with a relatively little
numerical effort. In particular, since the method is explicitly
designed for scale invariant systems, some deep connections with
CFT have been revealed \cite{mera-cft}.

\section{Entanglement entropy after a quantum quench}
\label{sec7}

The experimental realization of cold atomic systems with a high
degree of tunability of hamiltonian parameters, and the ability to
evolve in time with negligible dissipation, is motivating an
intensive study of extended quantum systems out of equilibrium.
New numerical algorithms have been developed to describe the
time-evolution of quantum systems effectively (among which
adaptive time-dependent DMRG \cite{tdmrg}, known as tDMRG has been
by now the most successful). As for the equilibrium counterpart,
the amount of entanglement of the time-dependent state governs the
effectiveness of the numerical methods based on MPS. It is then
fundamental to have firm bases and expectations for the
entanglement growth in non-equilibrium situations.  In this case
also conformal field theory has been a fundamental tool in
understanding very general properties of the time evolution of the
entanglement entropy.

The most studied situation (both theoretically and experimentally)
is a so called {\it quantum quench}. In a quench, an
extended quantum system is prepared at time $t=0$ in a pure state
$|\psi_0\rangle$ which is the ground state of some hamiltonian
$H_0$. For times $t>0$ the system is allowed to evolve {\em
unitarily} according to the dynamics given by a different
hamiltonian $H$, which may be related to $H_0$ by varying a
parameter such as an external field.

Based mainly on the first results from conformal field theory
\cite{cc-05,cc-07l} it has been possible to understand that the
entanglement entropy grows linearly with time for a so called
global quench (i.e. when the initial state differs globally from
the ground state and the excess of energy is extensive), while at
most logarithmically for a local one (i.e. when the initial
state has only a local difference with the ground state and so a
small excess of energy). As a consequence a local quench can
easily be simulated by means of tDMRG, while a global one is
harder and the numerics must be limited to relatively small
systems sizes. New time-dependent algorithms based explicitly on
the possibility of ``storing'' more entanglement
\cite{tMERA,h-07t,tmps} are being developed to have full access to
these dynamics.

\subsection{Global quench}

Suppose we prepare the system in a state $|\psi_0(x)\rangle$ and unitarily
evolve it with the hamiltonian $H$.
The matrix elements of the density matrix at time $t$ are
\begin{equation}
\fl
\langle\psi''(x'')|\rho(t)|\psi'(x')\rangle=
\langle\psi''(x'')|e^{-itH}|\psi_0(x)\rangle
\langle\psi_0(x)|e^{+itH}|\psi'(x')\rangle\,.
\end{equation}
We modify this time dependent expectation value as
\begin{equation}
\fl
\langle\psi''(x'')|\rho(t)|\psi'(x')\rangle=Z^{-1}
\langle\psi''(x'')|e^{-itH-\e H}|\psi_0(x)\rangle
\langle\psi_0(x)|e^{+itH-\e H}|\psi'(x')\rangle\,, \label{dm0}
\end{equation}
where we have included damping factors $e^{-\e H}$ in such a way
as to make the path integral absolutely convergent. We shall see
at the end of the calculation whether it is justified to remove
them. The normalization factor $Z=\langle\psi_0(x)|e^{-2\e
H}|\psi_0(x)\rangle$ guarantees that Tr\,$\rho=1$.

\begin{figure}[t]
\includegraphics[width=0.47\textwidth]{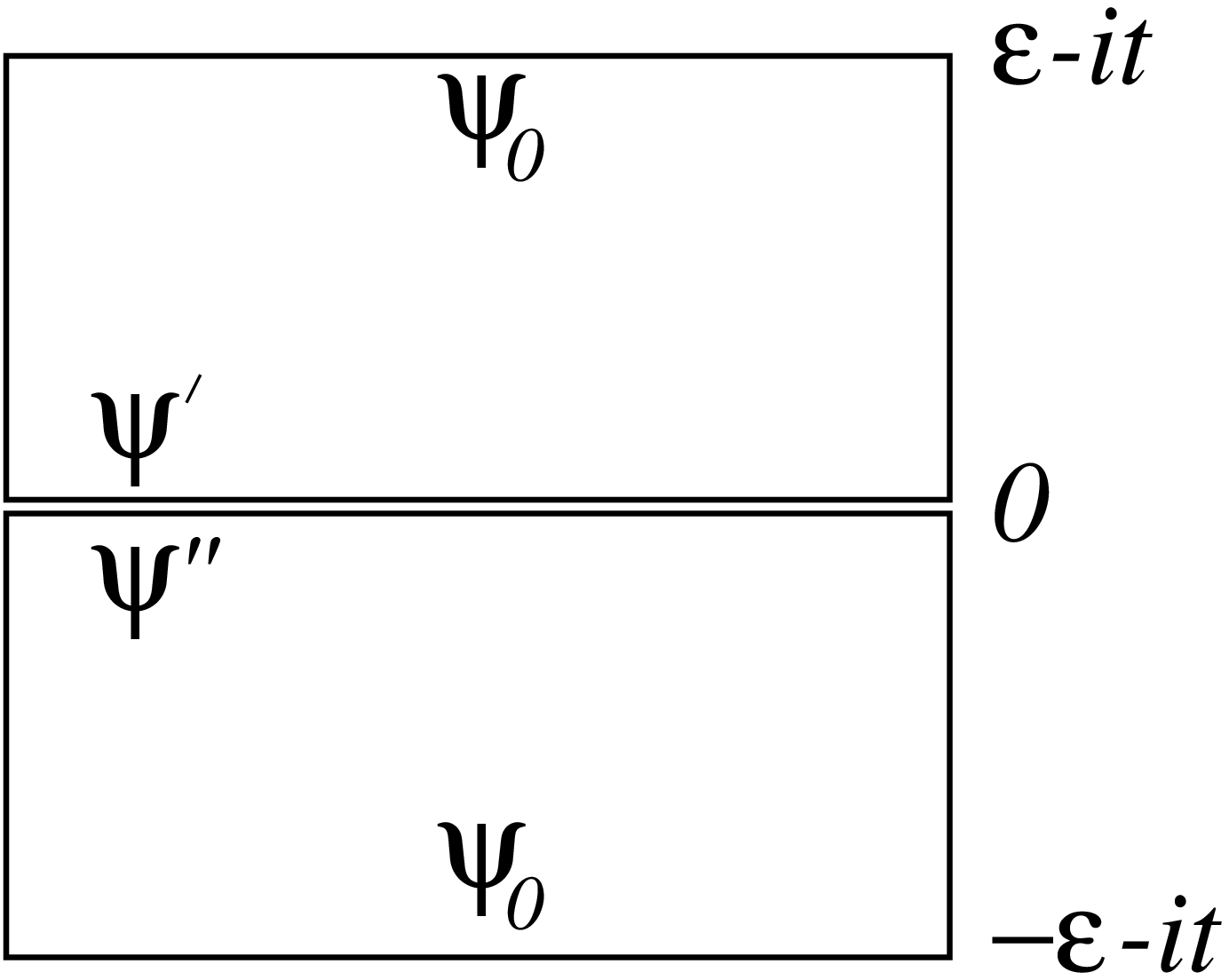}
\hspace{0.02\textwidth}
\includegraphics[width=0.48\textwidth]{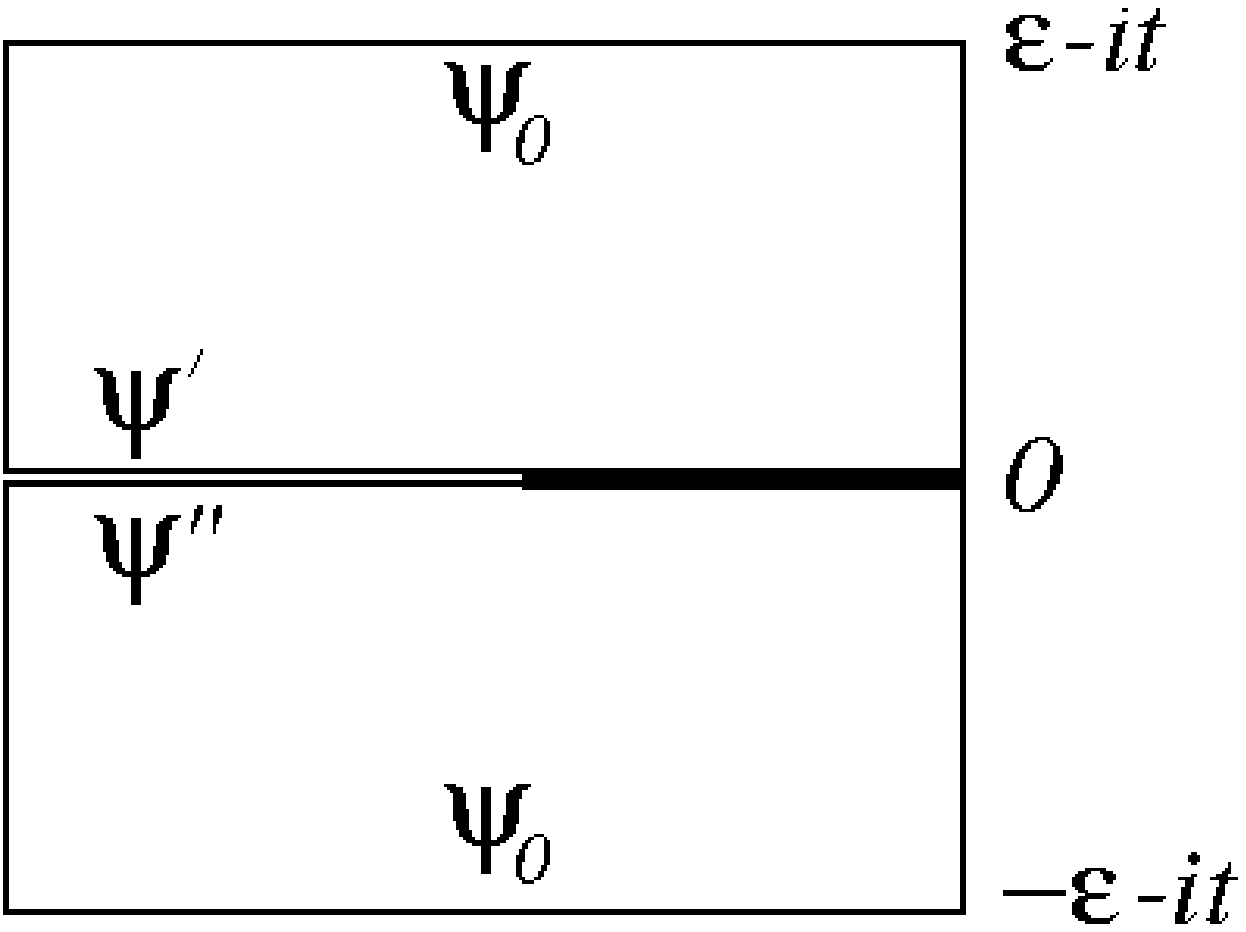}
\caption{Left:
Space-imaginary time regions for the density matrix in (\ref{dm0}).
Right: The reduced density matrix $\rho_A$ is obtained
by sewing together along $\tau=0$ only those parts of the $x$-axis
corresponding to points in $B$ (right part in this plot).}
\label{figpi}
\end{figure}

Each of the factors may be represented by an analytically
continued path integral in imaginary time: the first one over
fields $\psi(x,\tau)$ which take the boundary values $\psi_0(x)$
on $\tau=-\e-it$  and $\psi''(x)$ on $\tau=0$, and the second with
$\psi(z,\tau)$ taking the values $\psi'(x)$ on $\tau=0$ and
$\psi_0(x)$ on $\tau=\e-it$. This is illustrated in
Fig.~\ref{figpi}.
$Z$ is given by the euclidean path
integral over imaginary time $2\e$, with initial and final
conditions both equal to $\psi_0(x)$. This is the same as sewing
together the two edges in Fig.~\ref{figpi} along $\tau=0$.
As before, the reduced density matrix $\rho_A(t)$ is obtained
by sewing together along $\tau=0$ only those parts of the $x$-axis
corresponding to points in $B$, leaving open slits along $A$, and
Tr$\,\rho_A^n$ is given by sewing together $n$ copies of this in a
cyclic fashion. Thus
\begin{equation}
{\rm Tr}\,\rho_A^n=Z_n(A)/Z^n\,,
\end{equation}
where $Z_n$ is the path integral on an $n$-sheeted surface, where
the edges of each sheet correspond to imaginary times $-\tau_1$
and $\tau_2$, analytically continued to $\tau_1=\e+it$ and
$\tau_2=\e-it$, and the branch points lie along $\tau=0$ at the
boundaries points between $A$ and $B$. Finally, the entanglement
entropy is given by the derivative of ${\rm Tr}\,\rho_A^n$ with
respect to $n$ at $n=1$.

Eq.~(\ref{dm0}) has the form of the equilibrium expectation value
in a $1+1$-dimensional strip geometry with particular boundary conditions.
We wish to study this in the limit when $t$ and $\ell$ are much larger than
the microscopic length and time scales, when RG theory can be applied.
If $H$ is at (or close to) a quantum critical point, the bulk
properties of the critical theory are described by a bulk RG fixed
point (or some relevant perturbation thereof). In that case, the
boundary conditions flow to one of a number of possible
boundary fixed points \cite{dd}. Thus, for the purpose of
extracting the asymptotic behaviour, we may replace
$|\psi_0\rangle$ by the appropriate RG-invariant boundary state
$|\psi_0^*\rangle$ to which it flows. The difference may be taken
into account, to leading order, by assuming that the RG-invariant
boundary conditions are not imposed at $\tau=0$ and $\tau=2\e$ but
at $\tau=-\tau_0$ and $\tau=2\e+\tau_0$. In the language of
boundary critical behaviour, $\tau_0$ is called the extrapolation
length \cite{dd}. It characterizes the RG distance of the actual
boundary state from the RG-invariant one. It is always necessary
because scale-invariant boundary states are not in fact
normalizable \cite{cardy-05}. It is expected to be of the order of
the typical time-scale of the dynamics near the ground state of
$H_0$, that is the inverse gap $m_0^{-1}$. The effect of
introducing $\tau_0$ is simply to replace $\e$ by $\e+\tau_0$. The
limit $\e\to0^+$ can now safely be taken, so the width of the strip
is then taken to be $2\tau_0$.

\subsubsection{One interval in the infinite chain.}
Now we consider the case when $H$ is critical and the field theory
is a CFT. First let us consider the case when $A$ is a slit of
length $\ell$ and $B$ is the rest of the real axis. For real
$\tau$ the strip geometry described above may be obtained from the
upper half-plane by the conformal mapping $w=(2\tau_0/\pi)\log z$,
with the images of the slits lying along ${\rm arg}
z=\pi\tau_1/2\tau_0$. The result for $Z_n/Z^n$ in the upper half
$z$-plane, with two branch points follows from Eq.~(\ref{Fn})
where half of the points are the images with respect to the real
axis. To obtain the result in the strip geometry we transform this
two-point correlation function according to Eq.~(\ref{2pttra}).

After some algebra and continuing to $\tau_1=\tau_0+it$ (see
\cite{cc-05,cc-06,cc-07} for detailed calculations), we find for
$t,\ell\gg\tau_0$
\begin{equation}
\fl
{\rm Tr}\,\rho_A^n(t)\simeq c_n \left(\frac{\pi}{2\tau_0}\right)^{2d_n}
\left(\frac{e^{\pi\ell/2\tau_0}+e^{-\pi\ell/2\tau_0}+2\cosh(\pi t/\tau_0)}
{(e^{\pi\ell/4\tau_0}-e^{-\pi\ell/4\tau_0})^2 \cosh^2(\pi
t/2\tau_0)}\right)^{d_n} \t{\cal F}_n(x)\,.
\end{equation}
$ \t{\cal F}_n(x)$ is the boundary analogue of ${\cal F}_n(x)$ for
the four-point function of twist fields in the plane in
Eq.~(\ref{Fn}). After the conformal mapping and analytically
continuing the four-point ratio $x$ becomes for $\ell/\tau_0$ and
$t/\tau_0$ large \cite{cc-06,cc-07} \be x \sim \frac{e^{\pi
t/\tau_0}}{e^{\pi \ell/2\tau_0}+ e^{\pi t/\tau_0}}\,, \ee Thus for
$t<\ell/2$ we have $x\sim 0$ and in the opposite case $t>\ell/2$
we have $x\sim1$. Even if $\t{\cal F}_n(x)$ is generally unknown,
we only need its behaviour close to $x\sim 0$ and $1$, that are
easily deduced from general scaling. Indeed when $x\sim1$ the two
points are deep in the bulk, meaning $\t{\cal F}_n(1)=1$.
Oppositely when $x\ll 1$, the points are close to the boundary and
again $\t{\cal F}_n(0)=1$ (this because $\bra \tw\ket\neq0$, see
\cite{cc-07,cl-91}). Thus for the purpose of extracting the
asymptotic behaviour, the function $\t{\cal F}_n(x)$ is
irrelevant, explaining why the results of Ref.~\cite{cc-05},
obtained within this assumption are correct. Putting everything
together, in the case where $\ell/\tau_0$ and $t/\tau_0$ are
large, the moments of the reduced density matrix simplifies to
\begin{equation}
{\rm Tr}\,\rho_A^n(t)\simeq
c_n(\pi/2\tau_0)^{2d_n}\left(\frac{e^{\pi\ell/2\tau_0} +e^{\pi
t/\tau_0}} {e^{\pi\ell/2\tau_0}\cdot e^{\pi t/\tau_0}}\right)^{d_n}\,.
\end{equation}

Differentiating wrt $n$ to get the entropy, \be S_A(t)\simeq
-\frac{c}3 \log \tau_0+ \cases{ \displaystyle      \frac{\pi c
t}{6\tau_0}    \qquad t<\ell/2 \;,\cr \displaystyle      \frac{\pi
c\,\ell}{12\tau_0}\qquad t>\ell/2\,, } \label{SAt2} \ee that is
$S_A(t)$ increases linearly until it saturates at $t=\ell/2$. The
sharp cusp in this asymptotic result is rounded over a region
$|t-\ell/2|\sim\tau_0$. As a difference with Ref.~\cite{cc-05},
following \cite{sc-08}, we have added explicitly the subleading
constant term $\log\tau_0$ confirming that $\tau_0$ is connected
to the inverse mass gap in the initial state.

The result of the entanglement entropy for large time is the same
of a mixed state at inverse large finite temperature ${\beta}_{\rm
eff}=4\tau_0$ (see Eq.~(\ref{finiteT})). The physical
interpretation of this important effect is that any finite
subsystem $A$ reaches a quasi-stationary thermal state, in which
the infinite remaining part of the system $B$ act as a thermal
bath. It has been shown that, within CFT, this effective
temperature is the same for any observable \cite{cc-07} and so can
be properly defined. The possibility of defining a Gibbs-like
asymptotic state for a general hamiltonian governing the time
evolution (i.e. beyond the CFT case) is a subject of an intensive
current activity that would require its own review and that will
not be considered at all here.

These results for translationally invariant states have been
generalized to inhomogeneous quantum quenches with sharp
\cite{chl-08} and smooth \cite{sc-08} initial states.

\subsubsection{Physical interpretation.}
The qualitative, and many of the quantitative, features of $S_A(t)$
found above may be understood physically as follows \cite{cc-05}.
The initial state $|\psi_0\rangle$ has a very high energy relative to
the ground state of the hamiltonian $H$ which governs the
subsequent time
evolution, and therefore acts as a source of quasiparticle excitations.
Particles emitted from different points (further apart than the
correlation length in the initial state $\propto\tau_0$) are incoherent,
but pairs of particles moving to the left or right from a given point are
highly entangled. We suppose that the cross-section for producing such a pair
of particles of momenta $(p',p'')$ is $f(p',p'')$, and that, once they
separate, they move classically.
This will of course depends on $H$ and the state $|\psi_0\rangle$, and in
principle is calculable, but we made no strong assumptions on its form.
If the quasiparticle dispersion relation is $E=E_p$, the classical velocity
is $v_p=dE_p/dp$.
We assume that there is a maximum allowed speed which is taken to be 1, that is
$|v_p|\leq 1$. A quasiparticle of momentum $p$ produced at $x$ is
therefore at $x+v_pt$ at time $t$, ignoring scattering effects.

\begin{SCfigure}
\includegraphics[width=0.56\textwidth]{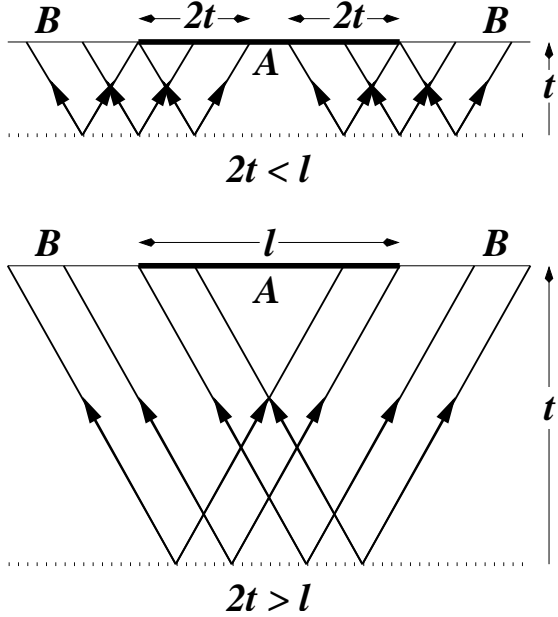}
\caption{
Space-time picture illustrating how the entanglement between
an interval $A$ and the rest of the system,
due to oppositely moving coherent quasiparticles, grows
linearly and then saturates. The case where the particles move only
along the light cones is shown here for clarity.
Reprinted with permission from \cite{cc-05}.}
\label{fig2tl}
\end{SCfigure}

Now consider these quasiparticles as they reach either $A$ or $B$ at
time $t$. The field at some point $x'\in A$ will be entangled with that
at a point $x''\in B$ if a pair of entangled particles emitted from a
point $x$ arrive simultaneously at $x'$ and $x''$ (see Fig.~\ref{fig2tl}).
The entanglement entropy between $x'$ and $x''$
is proportional to the length of the interval in $x$ for which this can
be satisfied. Thus the total entanglement entropy is
\be\fl
S_A(t)\approx \int_{x'\in A} \hspace{-3mm}
dx'\int_{x''\in B} \hspace{-3mm}  dx''\int_{-\infty}^\infty
dx\int
f(p',p'')dp'dp''\delta\big(x'-x-v_{p'}t\big)\delta\big(x''-x-v_{p''}t\big).
\ee
Now specialize to the case where $A$ is an interval of length $\ell$.
The total entanglement is twice that
between $A$ and the real axis to the right of $A$, which corresponds to
taking $p'<0$, $p''>0$ in the above. The integrations over the
coordinates then give ${\rm max}\,\big((v_{-p'}+v_{p''})t,\ell\big)$, so
that
\bea
\fl S_A(t)&\approx& 2t\int_{-\infty}^0dp'\int_0^\infty dp''f(p',p'')
(v_{-p'}+v_{p''})\,H(\ell-(v_{-p'}+v_{p''})t)+
\nonumber\\ \fl &+&
2\ell \int_{-\infty}^0dp'\int_0^\infty dp''f(p',p'')
\,H((v_{-p'}+v_{p''})t-\ell)\,,
\label{ppp}
\eea
where $H(x)=1$ if $x>0$ and zero otherwise. Now since $|v_p|\leq 1$,
the second term cannot contribute if $t<t^*=\ell/2$, so that $S_A(t)$ is
strictly proportional to $t$. On the other hand as $t\to\infty$, the
first term is negligible (this assumes that $v_p$ does not vanish
except at isolated points), and $S_A$ is asymptotically proportional to
$\ell$, as found earlier.

However, unless $|v|=1$ everywhere (as is the case for the CFT calculation),
$S_A$ is not strictly proportional
to $\ell$ for $t>t^*$.
The rate of approach depends on the behaviour of
$f(p',p'')$ in the regions where $v_{-p'}+v_{p''}\to 0$. This generally
happens at the zone boundary, and, for a non-critical quench, also at
$p'=p''=0$.
The exact form of $f(p',p'')$ has been exactly calculated only for the XY model
in a transverse field \cite{fc-08}.
The linear increasing followed by (almost) saturation has been checked
in several lattice models both analytically and numerically
\cite{cc-05,dmcf-06,eo-06,cdrz-07,pz-07,swvc-08,lk-08,fc-08,ep-08,mwnm-09,bkms-09},
but we do not have room here to discuss
the several interesting features that emerged like power-law behaviour
for large time, periodic time-oscillations etc.

It is worth mentioning that Eisler and Peschel \cite{ep-08} built
a lattice model with an exactly linear dispersion relation, and the
resulting time-dependent entanglement entropy is exactly the one calculated
within CFT.
It has been also argued that in random spin-chains the initial
linear growth of the entanglement entropy is replaced by a logarithmic
one \cite{bo-07}. This is a consequence of the strong scattering among
quasi-particles and seems to agree with numerical simulations \cite{dmcf-06}.

\subsubsection{General result for an arbitrary number of intervals.}
A general result can be also derived in the case when
$A$ consists of the union of the $N$
intervals $(u_{2j-1},u_{2j})$ where $1\leq j\leq N$ and
$u_k<u_{k+1}$. ${\rm Tr}\,\rho_A^n$ is given as usual by the ratio
$Z_n/Z^n$ which has the form of a correlation function
\begin{equation}
\label{cf12}
\langle\prod_j\tw_n(u_{2j-1}+i\tau_1) \prod_j\t\tw_n(u_{2j}+i\tau_1)\rangle\,,
\end{equation}
in a strip of width $2\tau_0$. We only need the asymptotic
behaviour of this correlation function for time $t$ and
separations $|u_j-u_k|$ much larger than $\tau_0$. Consequently
the complicated function ${\cal F}_{n,N}$ in Eq.~(\ref{generaln})
can be set to unity as before. After long algebra one arrives to
\cite{cc-05}
\begin{equation}
S_A(t)\sim S_A(\infty) +\frac{\pi c}{12\tau_0}
\sum_{k,l}(-1)^{k-l-1}{\rm max}(u_k-t,u_l+t)\,.
\end{equation}
If $N$ is finite (or more generally the $u_k$ are bounded) the
second term vanishes for sufficiently large $t$. At shorter times,
$S_A(t)$ exhibits piecewise linear behaviour in $t$ with cusps
whenever $2t=u_k-u_l$, at which the slope changes by $\pm\pi c/6\tau_0$
according to whether $k-l$ is even or odd. In the
case of an infinite number of regular intervals, with $u_k=k\ell$,
$k\in{\rm Z}$, $S_A(t)$ exhibits a sawtooth behaviour.

This behaviour can be explained in terms of the quasi-particles arguments of
the previous subsection, in which particles entering in and exiting from $A$
entangle and disentangle respectively, giving rise to the non-monotonic
behaviour.

The same reasoning applies to other situations, as for example the
time evolution in the presence of boundaries (particularly
relevant for tDMRG that are always performed with free boundary
conditions). In the simplest instance of the entanglement entropy
of the segment $[0,\ell]$ with the rest of the system, the
quasi-particle argument is easily understood for a perfect
reflecting wall at $x=0$, for which the resulting \em effective
\em length of the block is $2\ell$ and the saturation time
$t^*=\ell$, the double of periodic case. This is also easily
worked out from the conformal mapping $z=\sin(\pi w/2\tau_0)$
\cite{dmcf-06,cc-07}.

\begin{SCfigure}
\includegraphics[width=7cm]{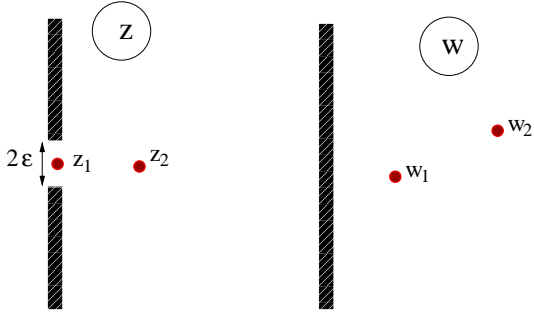}
\caption{Space-time region for the density matrix for the local
quench (left) mapped to the half-plane (right) by means of
Eq.~(\ref{mapp1}). $z_1=i\tau$ and $z_2=i\tau+\ell$ during the
computation and in the end $\tau\to i t$. Reprinted with
permission from \cite{cc-07l}.} \label{map1}
\end{SCfigure}

\subsection{Local quench}

Suppose we physically cut a spin chain at the boundaries between
two subsystems $A$ and $B$, and prepare a state where the individual
pieces are in their respective ground states. In this state the
two subsystems are unentangled, and its energy differs
from that of the ground state by only a finite amount. Let us join
up the pieces at time $-t$ and watch the system evolve up to
$t=0$. The procedure for the global quenches does not apply because the
initial state is not translational invariant and will not flow
under the renormalization group toward a conformally invariant
boundary state.

We can represent the corresponding density matrix in terms of path
integral on a modified world-sheet. The physical cut corresponds
to having a slit parallel to the (imaginary) time axis, starting
from $-\infty$ up to $\tau_1=-\e-i t$ (the time when the two
pieces have been joined), and analogously the other term of the
density matrix gives a slit from $\tau_2=\e-i t$ to $+\infty$,
like in Eq.~(\ref{dm0}). Again we introduced the regularization
factor $\e$. For computational simplicity we will consider the
translated geometry with two cuts starting at $\pm i\e$ and
operator inserted at imaginary time $\tau$. This should be
considered real during the course of all the computation, and only
at the end can be analytically continued to $it$. This plane with
the two slits is pictorially represented on the left of
Fig.~\ref{map1} where $i\tau$ corresponds to $z_1$. As shown in
the same figure, the $z$-plane is mapped into the half-plane ${\rm
Re}\, w>0$ by means of the conformal mapping \be
w=\frac{z}{\e}+\sqrt{\left(\frac{z}{\e}\right)^2+1} \qquad {\rm
with \,\, inverse} \qquad z=\e \frac{w^2-1}{2w}\,. \label{mapp1}
\ee On the two slits in the $z$ plane (and so on the imaginary
axis in the $w$ one) conformal boundary conditions compatible with
the initial state must be imposed.

We consider the time evolution of the entanglement entropy after
the local quench of two half-chains joined together at the point
$r_D=0$. We consider the four different spatial partitions of the
system depicted in Fig.~\ref{confs} among which we calculate the
entanglement.

\subsubsection{Case I: Entanglement of the two halves.}

We start with the more natural division, considering the
entanglement entropy between the two parts in which the system was
divided before the quench.
This is the case when $B$ is the positive real axis and $A$ is the negative
real axis.
$\Tr \rho_A^n$ transforms like a one-point function that in the
$w$ plane is $[2{\rm Re} w_1]^{-d_n}$. Thus in the $z$ plane at
the point $z_1=(0,i\tau)$ we have
\be
\langle\tw_n(z_1) \rangle=
\tilde{c}_n \left( \left|\frac{dw}{dz}\right|_{z_1}
\frac{a}{[2{\rm Re} w_1]}\right)^{d_n}
\ee
that using $\e w_1= i\tau+\sqrt{\e^2-\tau^2}$ becomes
\be
\langle\tw_n\rangle=\tilde{c}_n\left(\frac{a\e/2}{\e^2-\tau^2}\right)^{d_n}\,.
\label{phin}
\ee
Continuing this result to real time $\tau\to it$ we obtain
\be
\langle\tw_n(t)\rangle=\tilde{c}_n\left(\frac{a\e/2}{\e^2+t^2}\right)^{d_n}\,.
\ee
Using finally the replica trick to find the entanglement
entropy we have
\be
S_A=- \left.\frac{\partial }{\partial n}\Tr
\rho_A^n\right|_{n=1}= \frac{c}{6}\log \frac{t^2+\e^2}{a\e/2}
+\tilde{c}'_1\,.
\label{Shalf}
\ee
For $t\gg\e$ we have
\be
S_A(t\gg\e)=\frac{c}{3}\log \frac{t}a +k_0\,, \label{Slogt}
\ee
i.e. the leading long time behaviour is only determined by the
central charge of the theory in analogy with the ground state
value for a slit. This could result in a quite powerful tool to
extract the central charge in time-dependent numerical
simulations.
The constant $k_0$ is given by $k_0=\tilde{c}'_1+(c/6)\log(2a/\e)$.

\begin{figure}[t]
\epsfig{width=9cm,file=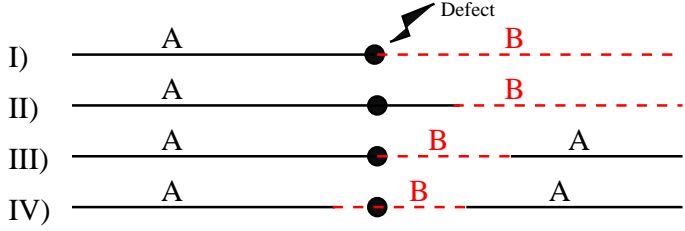} \caption{The four different
bipartitions of the line we consider in \cite{cc-07l}.
Reprinted with permission from \cite{cc-07l}.}
\label{confs}
\end{figure}

The behaviour for short time allows instead to fix the regulator
$\e$ in terms of the non-universal constant $\tilde{c}'_1$. In
fact imposing $ S_A(t=0)=0$ we have $\e= a/2
e^{-6\tilde{c}'_1/c}$. Consequently Eq.~(\ref{Shalf}) has no free
dynamical parameter, in contrast to the case of the global quench.

This short-time $\log t$ behaviour has been observed in different situations
with a local defect \cite{gkss-05,ep-07,ekpp-07,kkmgs-07,pv-08,pp-09,hgf-09}.

\subsubsection{Case II: De-centered defect.}

Let us now consider the entanglement of the region $r>\ell$ with
the rest of the system. In this case $\Tr \rho_A^n$ is equivalent
to the one-point function in the plane $z$ at the point
$z_2=\ell+i\tau$ as in Fig.~\ref{map1}. Using the conformal
mapping (\ref{map1}), analytically continuing and taking
$t,\ell\gg \e$ one finally get \cite{cc-07} \be S_A= \cases{
\frac{c}6 \log \frac{2\ell}a + \tilde{c}'_1  & $t<\ell$\,,\cr
\frac{c}6 \log \frac{t^2-\ell^2}{a^2} + k_0 & $t>\ell$\,, } \ee
with $k_0$ the same as in Eq.~(\ref{Slogt}). The interpretation of
this result is direct. Indeed at $t=0$ the joining procedure
produces a quasi-particle excitation at $r=0$ that propagates
freely with the corresponding speed of sound $v_s$ that in the CFT
normalization is $v_s=1$. This excitation takes a time $t=\ell$ to
arrive at the border between $A$ and $B$ and only at that time
will start modifying their entanglement. The following evolution
for $t\gg\ell$ is the same as in Eq.~(\ref{Slogt}).

Also the constant value for $t<\ell$ deserves a comment: it is
exactly the value known from CFT for the slit in the half-line
Eq.~(\ref{Sbou}). This is a non-trivial consistency check. Note
that a finite $\e$ smooths the crossover between the two regimes
and makes the entanglement entropy a continuous function of the
time.

\subsubsection{Cases III and IV: A finite slit.}

Let us consider again the same physical situation as before, but
we now calculate the entanglement entropy of $A=[0,\ell]$ and $B$
the remainder. For $t<0$ the real negative axis is decoupled from
the rest and does not contribute to the entanglement entropy, that
is just the one of a slit in half-chain.
The entanglement entropy is obtained from the replica trick
considering the scaling of a two-point function between the
endpoints of the slit, that must be mapped and analytically continued.
After long algebra one gets \cite{cc-07}
\be
{\rm Tr} \rho_A^n=
 \cases{
 \tilde{c}_n^2
 \left(\frac{a^2}{t^2}\frac{\ell+t}{\ell-t}\frac\e{4\ell}
 \right)^{x_n}\t{\cal F}_n\left(\frac{2t}{\ell+t}\right) &    $t<\ell$\,,\cr
 \tilde{c}_n^2\left(\frac{a^2}{\ell^2}\right)^{x_n} &
 $t>\ell$\,.
} \ee Note that compared with Ref.~\cite{cc-07l} we corrected the
behaviour for $t<\ell$ with  the generally unknown function
$\t{\cal F}_n(x)$, the boundary analogue of Eq.~(\ref{Fn}), and we
used $x=2t/(\ell+t)$ \cite{cc-07l} for the four-point ratio. For
$t>\ell$, only $\t{\cal F}_n(1)=1$ enters and the prediction of
Ref.~\cite{cc-07l} remains correct.
Using finally the replica trick,
we get the entanglement entropy
($\t{\cal F}'_1(x)=-\partial_n \t{\cal F}_n(x)|_{n=1}$)
\be\fl
S_A=\cases{
\frac{c}3 \ln \frac{t}a+\frac{c}6 \ln\frac\ell\e
+\frac{c}6 \ln 4\frac{\ell-t}{\ell+t}
+\t{\cal F}'_1\left(\frac{2t}{\ell+t}\right)+2\tilde{c}'_1&
 $t<\ell$\,, \label{plateau}\cr
\frac{c}3 \ln \frac{\ell}a +2\tilde{c}'_1& $t>\ell$\,.
}
\ee
The crossover time $t^*=\ell$ is again in agreement with the quasi-particles
interpretation.

There are several interesting features of this result. For very
short time $t\ll \ell$ it reduces to the $\ell=\infty$ case
Eq.~(\ref{Slogt}) as it should. The leading term for $t>\ell$ is
just the ground state value for a slit in an infinite line.
However the subleading term is not the same,
signaling that for long time the system still remembers something
of the initial configuration as a boundary term that is unable to
``dissipate''. Since the extra energy never dissipates under
unitary evolution, there is no reason for the constant terms to be
the same.  According to Eq.~(\ref{Bent}) these two constant terms
are the same only when $g=1$, as it is the case for the Ising
model with free boundary conditions.

Another interesting feature is the behaviour for $t<\ell$. This is
very similar to the form proposed in Ref.~\cite{ep-07} to fit the
numerical data, i.e.
\be S_A=\frac{c_0}3\log \ell
+\frac{c_1}{3}\log(t/\ell)+\frac{c_2}3\log(1-t/\ell)+k'\,.
\ee
Only the terms in $t+\ell$ and  $\t{\cal F}'_1$ were missing in
Ref.~\cite{ep-07}.
However these behaves smoothly for $0<t<\ell$ and their effect can be
well approximate by a constant factor in $k'$.
The results of the fit are $c_0\simeq 1+c_2$, $c_1\simeq 1$, and $c_2\simeq
1/2$ that are exactly our predictions for $c=1$.

For the most general case of a slit $A=[\ell_2,\ell_1]$, we remand
to the original paper \cite{cc-07l}. We stress here that the
result for short times (Eqs. (35) and (36) there) are generically
incorrect because we assumed $\t{\cal F}_n(x)=1$, that we learn
successively not to be the case. The result for $t>\ell_1$ are
instead correct, because in the regime only $\t{\cal F}(1)=1$
enters and in this case we have \be S_A(t>\ell_1)=\frac{c}3 \ln
\frac{\ell_1-\ell_2}a +2\tilde{c}'_1\,, \ee that is the same as
for case III. The correct results for $t<\ell_1$ in the various
regimes, can be read from Eqs. (48), (50) and (51) in
Ref.~\cite{cc-07l} from the general results of the two-point
functions of generic primary operators. We mention that we could
use non-equilibrium calculations/simulations to determine the
unknown function $\t{\cal F}_n(x)$. It is still not clear if this
could be effective.

An interesting generalization of the local quantum quench in
models with gradients has been provided in Ref.~\cite{eip-09},
where again all the entanglement evolution can be understood in
terms of the quasi-particle picture.

\subsubsection{Decoupled finite interval.}
\label{seccur}

A natural question arising is how the results we just derived
change when we introduce more than one defect in the line. It is
straightforward to have a path integral for the density matrix: we
only need to have pairs of slits for $-\infty$ to $-i\e$ and from
$i\e$ to $+i\infty$ everywhere there is a defect. However it
becomes prohibitively difficult to treat this case analytically.
In order to begin to understand the case when a finite interval
interval is initially decoupled, we consider the case when it lies
at the end of a half-line.

So, let us consider a semi-infinite chain in which the $A$
subsystem is the finite segment $(-\ell,0)$ and the $B$ is the
complement $(0,\infty)$ and with the initial defect at $r_D=0$.
The space-time geometry describing this
situation is like the one just considered, with a wall at $-\ell+iy$
($y$ real) that represents the boundary condition.

In these circumstances the inverse conformal mapping between the
$z$ plane and the half-plane can be worked out using the
Schwarz-Christoffel formula. After long algebra one obtains
\cite{cc-07l} \be z(w)=i\left(\frac{\ell}{\pi}\log (iw)+b
\frac{-iw-1}{-iw+1}\right)\,, \label{exmap} \ee with the parameter
$b$ related to $\ell$ and $\e$ in a non-algebraic way. (A slit in
the full line is closely related to this transformation, the last
piece is replaced by $(w^2-1)/(w^2+1)$.) Unfortunately the mapping
(\ref{exmap}) is not analytically invertible and its exact use is
limited to numerical calculations that do not help us, since we
need to perform an analytical continuation. However, even if not
completely justified, we can take the limit $\ell\gg\e$, before
the analytical continuation to real time obtaining \cite{cc-07l}
\be \fl w=-i\exp\left[ \frac{\pi i
\e}{2\ell}\left(\sqrt{z^2/\e^2+1}-z/\e \right)\right]\,,
\quad\Rightarrow\quad z=i\frac{\ell}{\pi}\log(iw)+
i\frac{\e^2}{\ell} \frac{\pi}{4} \frac{1}{\log(iw)}\,.
\label{apprmap} \ee This approximation is expected to fail for
$t\geq\ell$. It is easy to perform the mapping, continuing to real
time $\tau\to it$, and for $t\gg\e$ we have \cite{cc-07l} \be
\langle\tw_n(i\tau)\rangle=\tilde{c}_n
\left[\frac{\pi\e}{4\ell}\frac{1}{t\sin(\pi
t/2\ell)}\right]^{d_n}\,. \ee Clearly this cannot make sense when
the argument of the power-law becomes negative (i.e. for
$t>2\ell$), signaling the expected failure of Eq.~(\ref{apprmap}).
The time when this approximation fails cannot be understood from
this calculation, but only in the comparison with explicit results
in real time (or by the exact use of Eq.~(\ref{exmap})). Using the
replica trick, for the entanglement entropy we obtain \be
S_A=\frac{c}{6}\log\left(\frac{4\ell}{\pi\e}t\sin\frac{\pi
t}{2\ell}\right) +\tilde{c}'_1\,. \label{S2} \ee One is tempted to
assume that this result can be correct for $t<\ell$ and that for
larger time it saturates as suggested by the quasiparticle
interpretation. In the case of an initially decoupled slit of
length $\ell$ in an infinite chain Eq.~(\ref{S2}) is still valid
with the replacements $2\ell\to\ell$ and $c/6\to c/3$ as follows
from a simple analysis \cite{ekpp-07}. The validity of this equation
has been carefully tested for the XX chain in Ref.~\cite{ekpp-07},
finding very good agreement for all $t<\ell$, confirming the naive
expectation. In Ref.~\cite{isl-09} a more complicated kind of
defects has been investigated, and the results always agree with
Eq.~(\ref{S2}) when describing a conformal hamiltonian.

\section{Local quench, quantum noise and measuring the entanglement}
\label{sec8}

The entanglement entropy has been revealed to be a useful quantity
for a deep theoretical understanding of extended quantum systems,
especially in connection with criticality and topological order
(see the review by Fradkin in this volume \cite{f-rev}). However,
the final check and goal of any theory is the comparison with
experiments. The intrinsic non-local nature of the entanglement
entropy makes any attempt at an experimental measurement
difficult, if not impossible. Some bounds relating $S_A$ to
thermodynamic observables have been derived \cite{krs-06}, but
this is still far from being an operational measure.

However, there is one recent interesting proposal to measure the
entanglement entropy out of equilibrium, in the setup of the local
quench we have just described \cite{kl-08,kl-08b,kl-09}. The main
idea of Klich and Levitov is to relate the entanglement between
two half-chains to the distribution of the electrons passing
towards the contact between them. They considered (as we did
above) two semi-infinite chains (which are leads in actual
experiments) initially disconnected and then at some time $t_0$
joined together, allowing the passage of electrons (if the leads
are two Fermi seas, the quasiparticles of above are real
electrons). The transport at this {\it quantum point contact} is
described by the theory of quantum noise. This approach describes
the probability distribution of transmitted charge using the
generating function $\chi(\lambda)=\sum_{n=-\infty}^\infty
P_ne^{i\lambda n}$, where $P_n$ is the probability to transmit $n$
charges in total. The function $\chi(\lambda)$ can be written in
terms of cumulants $C_m$ \be \label{chi}
  \log\chi(\lambda)=\sum_{m=1}^\infty {(i\lambda)^mC_m\over m!}.
\ee
The fundamental point is that the constants $C_n$ are measurable quantities ($C_2$
is measured in routine experiments, and also $C_n$ up to $n=5$ have been
measured in more difficult experiments).

The main result of Ref.~\cite{kl-08} is to establish a  relation
between the cumulants $C_n$ and the entanglement entropy of the
two-halves \be S_A=\sum_{m>0} {\alpha_m\over  m !}C_{m} ,\quad
\alpha_m= \Big\{
\begin{array}{cc}
    (2\pi)^m |B_{m}|, &  $m$\,\,{\rm even} \\
  0,  &  $m$\,\,{\rm odd}
\end{array} ,\label{Enoise} 
\ee 
where $B_m$ are Bernoulli numbers. For quantum
noise generated in the contact switching on (at $t_0$) and off (at
$t_1$), the current fluctuations are gaussian ($C_{m\neq 2}=0$),
with a variance $C_2={1\over \pi^2}\log\frac{t_1-t_0}{\tau}$,
where $\tau$ is a short time cutoff set by the contact switching
rapidity. Combined with Eq.~(\ref{Enoise}) this gives entropy
$S_A\sim (1/3)\log |t_1-t_0|$. In Ref.~\cite{kl-08} this has been
put in direct relation with the standard formula $S_A=c/3
\log\ell$ ($c=1$ of free electrons). However, we have seen in the
previous section Eq.~(\ref{Slogt}), that $c/3 \log t$ is a key
feature of the local quench that comes from the specific
time-scale $\e$ whose analogous here is $\tau$. In
Ref.~\cite{kl-08} also the reaction of the system to a train of
pulses (i.e. periodic switching on and off of the contact) has
been considered. When two point-contacts are activated at the same time the 
response should be given by Eq.~(\ref{S2}). 

We finally stress that the previous analysis is valid for free electrons, and it is unclear
at present how the treatment must be properly modified in general to take
into account interactions to describe other universality classes.
A first calculation for the Luttinger liquid theory of the quantum Hall point 
contact showed that the measured noise is always logarithmic, with a prefactor
not given by the central charge, but by a filling $\nu$ dependent constant \cite{hgf-09}
\be
\chi(\lambda)= \exp\left[{-\frac{\lambda^2}2 \frac\nu{\pi^2} \log \frac{\Delta t}\tau}\right]\,,
\ee
for $\Delta t=t_1-t_0\gg\tau$.
Furthermore, it has been shown that for the Ising model  the noise at a point contact
is algebraic instead of logarithmic \cite{hgf-09}, suggesting that the relation between the full
counting statistics and the entanglement could not be structural.

\section*{Acknowledgments}
We are indebted with several collaborators that contribute to the
original works that are partially reviewed here: M. Campostrini,
O. Castro-Alvaredo, B. Doyon, M. Fagotti, R. Fazio, A. Lefevre, B.
Nienhuis, S. Sotiriadis, E. Tonni. Furthermore during the last
five years, we benefitted from very fruitful discussions with many
colleagues, among which we particularly thank I. Affleck, V. Alba,
L. Amico, J.-S. Caux, C. Castelnovo, J. Eisert, F. Essler, P.
Fendley, E. Fradkin, J. I. Latorre, P. Le Doussal, J. Moore, V.
Pasquier, R. Santachiara, K. Schoutens, G. Sierra, L. Tagliacozzo,
E. Vicari. This work was supported in part by EPSRC grants  EP/D050952/1.
PC benefitted from a travel grant from ESF (INSTANS activity).

\end{document}